\documentclass[letterpaper,twocolumn,10pt]{article}
\usepackage{usenix}

\usepackage{algorithmic}
\usepackage{algorithm}
\usepackage{amsmath}    
\usepackage{amssymb}    
\usepackage{amsfonts}   

\usepackage{subfigure}
\usepackage{multirow}
\usepackage[table]{xcolor}
\usepackage{colortbl}
\usepackage{pifont}
\usepackage{tabularx} 
\usepackage{array}
\usepackage{listings}
\usepackage{geometry}
\usepackage{setspace}
\usepackage{makecell, booktabs}
\usepackage[most]{tcolorbox}
\usepackage{inconsolata}
\usepackage{threeparttable}

\definecolor{gptcolor}{RGB}{230,240,255}
\definecolor{kimicolor}{RGB}{240,230,255}
\definecolor{dscolor}{RGB}{255,240,240}
\definecolor{grokcolor}{RGB}{240,255,240}
\definecolor{methodcolor}{RGB}{140,90,255}

\usepackage{bbding}

\lstdefinelanguage{json}{
    basicstyle=\footnotesize\ttfamily\setstretch{1.1}, 
    numbers=left,
    numberstyle=\tiny\color{gray},
    stepnumber=1,
    numbersep=8pt,
    showstringspaces=false,
    breaklines=true,
    frame=single,
    backgroundcolor=\color[RGB]{248,248,248},
    rulecolor=\color[gray]{0.8},
    captionpos=b,
    xleftmargin=10pt,
    xrightmargin=10pt,
    aboveskip=1em,   
    belowskip=1em,   
    literate=
     *{0}{{{\color{blue}0}}}{1}
      {1}{{{\color{blue}1}}}{1}
      {2}{{{\color{blue}2}}}{1}
      {3}{{{\color{blue}3}}}{1}
      {4}{{{\color{blue}4}}}{1}
      {5}{{{\color{blue}5}}}{1}
      {6}{{{\color{blue}6}}}{1}
      {7}{{{\color{blue}7}}}{1}
      {8}{{{\color{blue}8}}}{1}
      {9}{{{\color{blue}9}}}{1}
      {:}{{{\color{black}:{}}}}{1}
      {,}{{{\color{black},}}}{1}
      {"}{{{\color{orange}"}}}{1},
}

\lstdefinestyle{tightjson}{
  basicstyle=\scriptsize\ttfamily\setstretch{0.9},  
  numbers=none,                                   
  frame=single,                                   
  backgroundcolor=\color[RGB]{248,248,248},        
  breaklines=true,                                 
  showstringspaces=false,                          
  captionpos=b,                                    
  xleftmargin=5pt,                                 
  xrightmargin=5pt,                                
  aboveskip=0.5em,                                 
  belowskip=0.5em                                  
}

\newenvironment{packeditemize}{
	\begin{list}{$\bullet$}{
			\setlength{\labelwidth}{4pt}
			\setlength{\itemsep}{0pt}
			\setlength{\leftmargin}{\labelwidth}
			\addtolength{\leftmargin}{\labelsep}
			\setlength{\parindent}{0pt}
			\setlength{\listparindent}{\parindent}
			\setlength{\parsep}{0pt}
			\setlength{\topsep}{1pt}}}{\end{list}}

\newcolumntype{Y}{>{\centering\arraybackslash}X}

\usepackage{graphicx} 
\usepackage{float}    
\usepackage{hyperref}

\usepackage{siunitx}
\begin{document}
\title{\textsc{ViPer} Strike: Defeating Visual Reasoning CAPTCHAs via \\Structured Vision–Language Inference\thanks{Accepted by \textcolor{violet}{USENIX Security'26}; \Envelope~for corresponding author.}}

\author{
{\rm Minfeng Qi$^\dag$, Dongyang He$^\dag$, Qin Wang$^{\flat,}\textsuperscript{\,\Envelope}$, Lefeng Zhang$^\dag$}\\
$^\dag$ City University of Macau\\
$^\flat$ CSIRO Data61\\
}

\maketitle
\thispagestyle{empty}
\author{}

\begin{abstract}



Visual Reasoning CAPTCHAs (VRCs) combine visual scenes with natural-language queries that demand compositional inference over objects, attributes, and spatial relations. They are increasingly deployed as a primary defense against automated bots. Existing solvers fall into two paradigms: \emph{vision-centric}, which rely on template-specific detectors but fail on novel layouts, and \emph{reasoning-centric}, which leverage LLMs but struggle with fine-grained visual perception. Both lack the generality needed to handle heterogeneous VRC deployments. 

We present \textsc{ViPer}, a unified attack framework that integrates structured multi-object visual perception with adaptive LLM-based reasoning. \textsc{ViPer} parses visual layouts, grounds attributes to question semantics, and infers target coordinates within a modular pipeline. Evaluated on six major VRC providers (VTT, Geetest, NetEase, Dingxiang, Shumei, Xiaodun), \textsc{ViPer} achieves up to 93.2\% success, approaching human-level performance across multiple benchmarks. Compared to prior solvers, GraphNet (83.2\%), Oedipus (65.8\%), and the Holistic approach (89.5\%), \textsc{ViPer} consistently outperforms all baselines. The framework further maintains robustness across alternative LLM backbones (GPT, Grok, DeepSeek, Kimi), sustaining accuracy above 90\%.

To anticipate defense, we further introduce \textit{Template-Space Randomization} (TSR), a lightweight strategy that perturbs linguistic templates without altering task semantics. TSR measurably reduces solver (i.e., attacker) performance. Our proposed design suggests directions for human-solvable but machine-resistant CAPTCHAs.

\end{abstract}

\section{Introduction}
Human Interaction Proofs (HIPs) have long served as a critical defense against automated abuse on the Internet. Over the past two decades, CAPTCHA (\textit{Completely Automated Public Turing test to tell Computers and Humans Apart}) mechanisms have evolved from distorted text recognition to increasingly complex visual challenges, in an arms race against rapidly advancing machine learning models~\cite{uzun2018rtcaptcha,tang2018captcha,shi2020text,guerar2021gotta,wang2023experimental,ousat2024matter}.  

Against this backdrop, a new class of \emph{Visual Reasoning CAPTCHAs} (VRCs) has emerged as one of the most challenging challenge-based defenses in practice. VRCs present users with visual scenes paired with natural-language instructions that require multi-step reasoning over spatial, comparative, and semantic attributes~\cite{wang2018captcha,gao2021research,wang2023extended,ding2025illusioncaptcha}. They are widely deployed by major providers such as Tencent Waterproof Wall~\cite{vtt} and Geetest~\cite{geetest}, protecting large-scale online services and processing millions of challenges daily~\cite{builtwith_recaptcha}.

\begin{figure}[!t]
    \centering
    \includegraphics[width=1.0\linewidth]{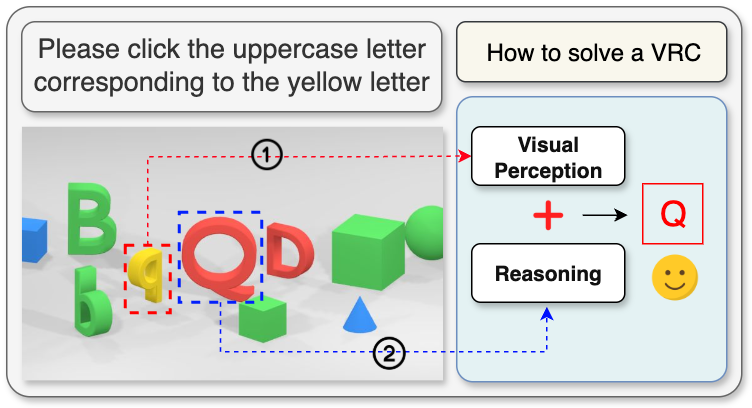}
    \vspace{-0.2in}
    \caption{\textbf{Example of a Visual Reasoning CAPTCHA (VRC).} A solver
    \ding{172} identifies the yellow lowercase letter (\emph{perception}); \ding{173} matches it with its corresponding uppercase form (\emph{reasoning}), producing the correct answer `Q'.}
    \label{fig:vrc_example}
    \vspace{-0.2in}
\end{figure}

\noindent\textbf{VRCs pose great challenges.}  
Unlike prior CAPTCHAs that primarily test isolated perceptual skills, VRCs jointly require two of the hardest open problems in modern AI: \textit{fine-grained visual perception} and \textit{language-guided compositional reasoning}. A solver must not only localize and distinguish multiple objects in cluttered scenes, but also interpret multi-constraint instructions expressed in natural language and bind them correctly to visual evidence. As illustrated in Fig.~\ref{fig:vrc_example}, the solver must first \ding{172} identify the yellow lowercase letter and then \ding{173} reason about a semantic transformation (``\textit{match to the uppercase form}'') to derive the correct answer `Q'. VRC providers further randomize object categories, layouts, and phrasing patterns~\cite{gao2021research,guerar2021gotta}, rendering per-template tuning ineffective.

\smallskip
\noindent\textbf{Why VRCs still matter.}  
In parallel, behavior-based systems such as reCAPTCHA~\cite{google2018recaptcha} have become dominant in Internet-wide deployments, covering approximately 88\% of websites in practice~\cite{builtwith_recaptcha}. However, prevalence does not imply robustness. A growing body of prior work has shown that reCAPTCHA-style mechanisms are increasingly vulnerable to learning-based attacks, with reported success rates exceeding 70\% under realistic threat models~\cite{george2017generative,plesner2024breaking,teohcaptchas}.  

This contrast is also evident in our own measurements. In a zero-shot evaluation of GPT-4o across widely used CAPTCHA types (Appendix Table~\ref{tab:captcha_types_summary}), behavior-centric challenges such as reCAPTCHA are solved with success rates approaching 50\%, whereas VRCs remain substantially more resistant, with accuracy dropping to only 31\%. These results highlight that VRCs occupy a harder frontier than many widely deployed alternatives, making them a critical target for understanding the true limits of automated visual–language reasoning and the future of challenge-based defenses.

\smallskip
\noindent\textbf{Existing paradigms are limited.}  
Early machine-learning (ML) attacks targeted VRCs with modular pipelines. The Holistic model~\cite{gao2021research} combined a BiLSTM network with rule-based reasoning and reported up to 88\% success on Tencent VTT~\cite{vtt}, later extended with improved template matching~\cite{wang2023extended}. More recently, VRC-GraphNet~\cite{xu2023vrc} uses nodes for objects and edges for relations, solved via a graph neural network. While effective on their target platforms, both approaches exhibit fundamental limitations: they encode only shallow heuristics rather than genuine reasoning, and thus fail to capture compositional semantics in queries.

The rise of large language models (LLMs) such as GPT-4o and Grok-2~\cite{achiam2023gpt,grok} challenges the hard AI problem. These models combine high-capacity reasoning with visual perception, and excel on benchmarks in visual question answering and chart interpretation~\cite{xu2023multimodal,zhou2024large,yu2025survey,chang2024survey,yang2024harnessing}. In principle, such capabilities should transfer naturally to VRCs. Indeed, Oedipus~\cite{deng2024Oedipus} was the first framework to leverage LLMs for solving VRCs by decomposing queries into symbolic sub-tasks for chain-of-thought reasoning. However, despite this innovation, its performance remains limited: on Tencent VTT it achieves only an average success rate of 63.5\%, far below human accuracy. The reliance on LLM-native encoders without robust visual grounding leaves Oedipus vulnerable to occlusion, cluttered scenes, and fine-grained attribute distinctions.

Neither paradigm provides both \textit{robustness} and \textit{generality}. ML-based solvers tend to overfit to surface features and break down under varying query phrasing and heterogeneous VRCs (cf. Appendix~\ref{sec:data_sources}). LLM-based solvers, in contrast, lack precise visual grounding and often misinterpret fine-grained spatial or attribute details. Bridging these shortcomings requires an integrated framework that combines reliable visual perception with structured reasoning to operate consistently across various VRCs.

\smallskip
\noindent\textbf{Our approach.}
We introduce \textsc{ViPer} (\textit{\textbf{V}isual \textbf{I}nference via \textbf{P}erception and \textbf{E}mbedded \textbf{R}easoning}), a general attack framework that \emph{decouples} structured visual perception from symbolic reasoning, reconnecting them through adaptive prompts. \textsc{ViPer} employs a high-precision multi-object detector to inventory candidate objects (shape, color, orientation, coordinates) from the VRC image. A lightweight parser extracts semantic constraints from the query, explicitly encoding both attribute filters and spatial relations. This structured representation is then processed by an LLM guided by adaptive prompts tailored to the reasoning type. Across six major VRC providers, \textsc{ViPer} achieves up to 93.2\% accuracy and consistently outperforms prior automated solvers, approaching human-level performance on several benchmarks while maintaining strong robustness across heterogeneous VRCs.

We make the following contributions:

\begin{packeditemize}

\item[$\triangleright$] \textit{A unified framework for breaking VRCs} (\S\ref{sec:method}).  
We introduce \textsc{ViPer}, a modular vision–language framework that casts VRC solving as a compositional \emph{perception–reasoning} task. By combining multi-object detection with prompt-adaptive LLM reasoning, \textsc{ViPer} generalizes across heterogeneous VRC templates without per-platform retraining.

\item[$\triangleright$] \textit{A multi-platform benchmark} (\S\ref{sec:benchmark}).  
We curate a benchmark of 6{,}000 VRC challenges across six major platforms, each annotated with ground-truth answers and object bounding boxes. In addition, we release a companion dataset of 1{,}200 multi-object annotated VRC scenes to train and assess perception modules. Together, these resources provide the first standardized basis for cross-platform comparison and reproducible development of solver architectures.

\item[$\triangleright$] \textit{Comprehensive evaluation} (\S\ref{sec:evaluation}).  
We evaluate \textsc{ViPer} under four dimensions: component ablation, baseline comparison, LLM backend swap, and response time. Across all settings, our framework achieves state-of-the-art performance, demonstrating superior accuracy and robustness compared to existing solvers.

\item[$\triangleright$] \textit{Defense approach.} (\S\ref{sec-defense}).  
We introduce Template-Space Randomization (TSR), a lightweight defense that perturbs questions phrasing through synonym substitution, relation rewording, and indirection. Preliminary experiments on 600 customized questions show consistent success-rate reductions, highlighting the brittleness of current language–vision alignments.

\end{packeditemize}

\smallskip
\noindent\textbf{Why we do this work.}  
Behavior-based systems (e.g., reCAPTCHA, VRCs) were widely deployed and protect high-volume services. However, we show that these systems represent a harder frontier than many alternatives, yet lack systematic security evaluation. Our work can document current limitations, guide vendors toward more robust designs, and provide a clear ethical reference point for human–machine differentiation mechanisms.

\section{Technical Warm-Ups}
\label{sec-rw}

\subsection{LLM-based Inference}

LLMs have strong capabilities in understanding, generating, and reasoning over natural language. Recent models (e.g., GPT-4o, Gemini-2.5) perform tasks requiring commonsense reasoning, instruction following, and spatial language comprehension with near-human proficiency.

Formally, let $\mathcal{Q}$ denote the space of natural-language queries and $\mathcal{A}$ the space of possible answers (e.g., textual labels, numerical values, or coordinates). An LLM can be represented as a mapping:
\(
f_{\mathrm{LLM}} : \mathcal{Q} \rightarrow \mathcal{A},
\)
where the model estimates:
\setlength{\abovedisplayskip}{2pt}
\setlength{\belowdisplayskip}{2pt}
\begin{equation}
f_{\mathrm{LLM}}(q) = \arg\max_{a \in \mathcal{A}} P(a \mid q; \theta),
\end{equation}
with $\theta$ denoting pre-trained model parameters fixed during inference. In practice, $f_{\mathrm{LLM}}$ is implemented via auto-regressive decoding from a prompt $q$, optionally augmented with system-level instructions.

Multimodal extensions of LLMs generalize this mapping to jointly process language and visual inputs. Let $\mathcal{I}$ denote the space of image inputs (e.g., raw pixels or symbolic descriptions). A multimodal LLM defines:
\setlength{\abovedisplayskip}{2pt}
\setlength{\belowdisplayskip}{2pt}
\begin{equation}
\label{eq-llm}
f_{\mathrm{LLM}} : \mathcal{Q} \times \mathcal{I} \rightarrow \mathcal{A},
\end{equation}
allowing the model to perform reasoning over paired image–text inputs. When visual information is pre-processed into a structured, symbolic representation by an external vision module, the problem reduces to language-only reasoning over high-level visual facts, enabling the LLM to leverage its pretrained semantic and logical priors without requiring vision-specific retraining.

LLMs also support diverse reasoning modes:

\begin{packeditemize}
    \item \textit{Attribute-based filtering:} selecting objects by specified properties such as color, shape, or textual label.
    \item \textit{Spatial and comparative reasoning:} interpreting positional or relational descriptions (e.g., ``\textit{to the left of}'', ``\textit{above the red square}'', ``\textit{largest among}'').
    \item \textit{Logical inference:} applying Boolean compositions, negation, ordering, or multi-condition constraints.
\end{packeditemize}

\begin{figure*}[!t]
    \centering
    \includegraphics[width=1.0\linewidth]{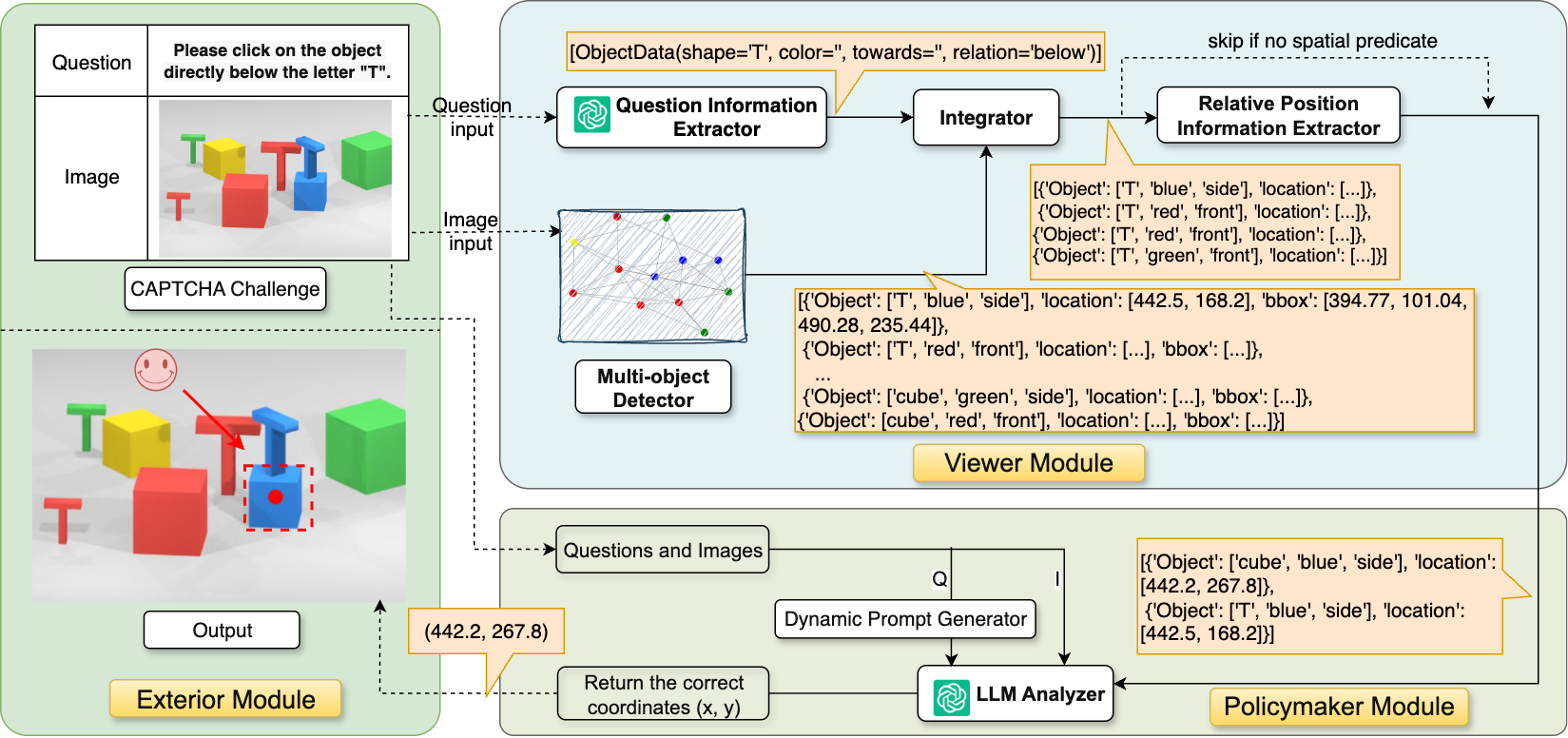}
    \vspace{-15pt}
    \caption{\textbf{\textsc{ViPer} end-to-end pipeline.} \textit{Exterior} ingests image $I$ and instruction $q$. \textit{Viewer} conducts structured perception: a multi-object detector yields labeled boxes $\mathcal{V}$; QIE parses $q$ into attribute slots; the Integrator aligns slots with $\mathcal{V}$ to obtain candidate set $V'$; if $q$ includes a spatial phrase, RPIE resolves references via geometric projection (\S\ref{sec:viewer-details}- \S\ref{sec:rpie}). \textit{Policymaker} composes a task-conditioned prompt from $q$ and $V'$, queries an LLM, and outputs a coordinate $(x,y)$ (\S\ref{sec:policymaker}).}
    \label{fig:final_VIPER}
    \vspace{-0.1in}
\end{figure*}

\subsection{Multi-object Detection}  
A multi-object detector serves as the perception backbone in our framework. It formulates detection as a regression problem from image pixels to bounding box coordinates and class probabilities. Given an input image $I \in \mathbb{R}^{H \times W \times 3}$, the detector partitions the image into an $S \times S$ grid and predicts $B$ candidate boxes per cell, each parameterized by $(x, y, w, h)$ for box geometry, an objectness score $C$, and class probabilities over $C_{\text{cls}}$ categories. The resulting output tensor has dimension $S \times S \times B \times (5 + C_{\text{cls}})$, where the $5$ encodes $(x, y, w, h, C)$. Coordinates $(x, y)$ are normalized to the grid cell and $(w, h)$ to the entire image.

Training is guided by a compound loss:
\setlength{\abovedisplayskip}{2pt}
\setlength{\belowdisplayskip}{2pt}
\begin{equation}
\mathcal{L} = \lambda_{\text{loc}}\mathcal{L}_{\text{bbox}} + 
\lambda_{\text{obj}}\mathcal{L}_{\text{obj}} + 
\lambda_{\text{cls}}\mathcal{L}_{\text{cls}},
\end{equation}
where $\mathcal{L}_{\text{bbox}}$ penalizes bounding box regression errors, $\mathcal{L}_{\text{obj}}$ measures objectness quality, and $\mathcal{L}_{\text{cls}}$ is the categorical classification loss. Hyperparameters $(\lambda_{\text{loc}}, \lambda_{\text{obj}}, \lambda_{\text{cls}})$ balance these terms.

At inference, the detector outputs candidate boxes with confidence scores. Non-maximum suppression (NMS) filters redundant predictions, yielding the final set:
\setlength{\abovedisplayskip}{2pt}
\setlength{\belowdisplayskip}{2pt}
\begin{equation}
\label{eq-multi}
\mathcal{D} = \{(c_k, s_k, b_k)\}_{k=1}^K,
\end{equation}
where $c_k$ is the predicted class, $s_k$ the confidence score, and $b_k=(x_1,y_1,x_2,y_2)$ the bounding box in image space.

\section{Threat Model}
\label{sec-threat-model}

\noindent\textbf{Attacker} (VRC solver).
We assume that an adversary is an external entity with no privileged access to the target service’s internal infrastructure, proprietary CAPTCHA generation logic, or private training data. Interaction occurs solely through publicly accessible interfaces, such as web-based user authentication portals, under the same network and session constraints imposed on legitimate users. The goal is to solve VRC challenges at a rate that undermines their security purpose, automating what is intended to be a human-exclusive verification step.

However, an adversary can collect an arbitrary number of VRC instances available within normal usage limits, capture all rendered visual content and accompanying natural-language prompts, and process these using any computational resources under their control. This includes applying pre-trained LMMs, integrating specialized perception modules, and employing symbolic or programmatic reasoning pipelines. The model of interaction is strictly black-box: the attacker receives binary feedback on submission correctness but cannot obtain intermediate model states, gradients, or any direct control over challenge construction.

We exclude the exploitation of software vulnerabilities, backend compromise, or side-channel data exfiltration that are out of the scope of \textsc{ViPer}.

\smallskip
\noindent\textbf{Defender} (VRC provider).
We model the defender as the service operator who deploys VRCs to safeguard against automated abuse. The defender controls challenge generation and presentation, enforces rate limits, and applies server-side verification. All challenge content, both images and instructions, is assumed visible to users, and thus to client-side adversaries. The defender may choose from available VRC templates.  

The defender is to prevent automated solvers from achieving human-level success rates without manual effort. Our study evaluates how well current VRC deployments satisfy this goal under the adversarial case.

\section{Our \textsc{ViPer}}
\label{sec:method}

We present \textsc{ViPer}, a modular framework (Fig.\ref{fig:final_VIPER}) that takes as input a VRC image $I$ and its natural-language prompt $q$ and returns a click coordinate $(x,y)$ in image space. Our design explicitly decouples perception from reasoning: the vision module extracts structured objects and attributes, and a query-aware interface passes only task-relevant representations to the LLM. 

\subsection{System Architecture}\label{subsec:method-arc}

\textsc{ViPer} consists of three key modules. 

\smallskip
\noindent\underline{\textbf{\ding{182} Exterior.}} Exterior mediates interaction with the challenge environment. It receives the rendered image and associated question, normalizes inputs to the internal data schema, and formats the final answer for submission. No solver logic is implemented in this layer.

\smallskip
\noindent\underline{\textbf{\ding{183} Viewer.}} Viewer forms the perceptual backbone. It converts the raw image into a structured inventory of objects and attributes and aligns this inventory with the semantic intent of the prompt. A trained \textbf{\textit{multi-object detector}} (\S\ref{sec:viewer-details}) returns a collection of labeled hypotheses, where each hypothesis contains a \emph{compound} semantic label (shape–color–orientation from the ontology in Table~\ref{tab:vrc_labels}), a confidence score, and a pixel-space bounding box with its center; a representative serialization appears in Listing~\ref{lst:yolo_output}. 

The detector output is then filtered and (when necessary) augmented in two steps. 
\begin{packeditemize}
    \item First (\S\ref{sec:selector}), a Question Information Extractor \textbf{\textit{(QIE)}} parses $q$ into a structured query that enumerates the attributes and categories referenced by the instruction. Then, an \textbf{\textit{Integrator}} matches the QIE constraints against $\mathcal{D}$ to retain only semantically relevant detections; unspecified attributes in the query act as wildcards, and coarse categories such as ``\textit{letter}'', ``\textit{number}'', or ``\textit{3D object}'' are resolved via predefined type maps. 
    \item Second (\S\ref{sec:rpie}), when the prompt defines the target purely through a spatial relation to a reference (for example, ``\textit{the item to the left of the small gray cube}''), a Relative Position Information Extractor \textbf{\textit{(RPIE)}} infers candidate targets by projecting from the reference toward the indicated direction and testing containment against detected boxes. The result is a compact set of candidate objects $V'$, each represented by a normalized attribute tuple, bounding box, and center coordinate.
\end{packeditemize}




\smallskip
\noindent\underline{\textbf{\ding{184} Policymaker.}} Policymaker is the reasoning engine (\S\ref{sec:policymaker}). It classifies the reasoning type suggested by the parsed query (e.g., spatial, comparative, or attribute matching), constructs a dynamic prompt that includes the original instruction together with the serialized candidate set from the Viewer, and invokes a large language model to select the target. The output is constrained to a coordinate $(x,y)$  in the image. To reduce ambiguity, the prompt states the expected reasoning mode and provides only the attributes required to resolve the instruction.

\subsection{Structured Perception}
\label{sec:viewer-details}
The perception backbone is a single multi-object detector configured to emit \emph{compound} class labels that directly align with attributes referenced in VRC prompts. Each label encodes the triplet \textsc{shape}$\times$\textsc{color}$\times$\textsc{orientation}, ensuring that every detection carries the attributes most frequently queried in natural-language instructions. The ontology comprises $91$ categories spanning alphanumeric symbols and geometric primitives with five colors and two canonical poses. Encoding attributes at the label level avoids post hoc fusion of separate predictions and provides a stable interface for downstream reasoning.

\begin{table}[t]
    \centering
    \footnotesize
    \renewcommand{\arraystretch}{1}
    \caption{\textbf{Ontology of compound label attributes.} \emph{Shape}, \emph{Toward}, and \emph{Color} are derived from six VRC providers.}
    \label{tab:vrc_labels}
    \vspace{3pt}
    \begin{tabularx}{\columnwidth}{>{\columncolor{kimicolor}}cX}
        \toprule
        \multicolumn{1}{c}{\textbf{Label}} &  \multicolumn{1}{c}{\textbf{Label values}} \\
        \midrule
        Shape & 
        0–9, a–z, A–Z, sphere, cube, cone, cylinder, hexagonal prism, polyhedron, rectangular prism, slant, square pyramid, torus, triangular prism, triangular pyramid, square, parallelogram, round, rectangle, hexagon, trapezoidal, lozenge, triangle, pentagram, pentagon \\
        \midrule
        Toward & front, side \\
        \midrule
        Color & yellow, green, gray, blue, red \\
        \bottomrule
    \end{tabularx}
    \vspace{-0.15in}
\end{table}

\smallskip
\noindent\textbf{Training setup.}
To enable cross-platform generalization, we aggregate annotated images from six providers into a unified training corpus and normalize heterogeneous schemas by mapping platform-specific tags into a compound-label space (Appendix~\ref{sec:data_sources}). All images are resized to $640{\times}640$ with aspect-preserving padding. The detector is trained with a compound loss jointly optimizing localization, objectness, and classification, and uses standard data augmentations (mosaic/flipping/colour jittering) to improve robustness across visual styles.

\smallskip
\noindent\textbf{Inference output.}
At inference time, the network produces, for each image, a set of hypotheses (Equation~\ref{eq-multi})
\[
\mathcal{V}=\{(\ell_k, s_k, b_k)\}_{k=1}^{K},
\]
where $\ell_k$ is the compound label, $s_k\in[0,1]$ is the confidence, and $b_k=(x_1,y_1,x_2,y_2)$ are bounding-box corners in pixel space. For each retained detection, the center
\(
c_k=\big(\tfrac{x_1+x_2}{2},\tfrac{y_1+y_2}{2}\big)
\)
is computed and stored together with $(\ell_k,s_k,b_k)$. The Viewer serializes this structured output in a minimal record format and passes it to the Integrator for candidate pruning and, where applicable, later alignment with relational cues. A real and representative serialization case appears in Listing~\ref{lst:yolo_output}.

\begin{lstlisting}[style=tightjson, language=json,caption={\textbf{Serialized detection output for a VRC instance}, including compound label (Object), pixel-space center (location), and bounding box (bbox).}, label={lst:yolo_output}]
[
  {"Object": "cylinder", 
   "location": [366.9, 165.7], 
   "bbox": [335.7237854003906, 127.53125762939453, 398.072265625, 203.88307189941406]},
  {"Object": "6", 
   "location": [434.3, 171.2], 
   "bbox": [401.0405578613281, 118.474853515625, 467.6160888671875, 223.9849090576172]}
]
\end{lstlisting}

\subsection{Semantic Extraction and Alignment}
\label{sec:selector}

This stage links the natural-language instruction to the objects detected by the Viewer. The instruction is first parsed into a structured query, which is then aligned with the detector output to filter out irrelevant objects. When the target is defined only through its relation to a reference, this stage also encodes the necessary information for resolving the implied target in the next step.

\begin{figure}[!t]
    \centering
    \includegraphics[width=1.0\linewidth]{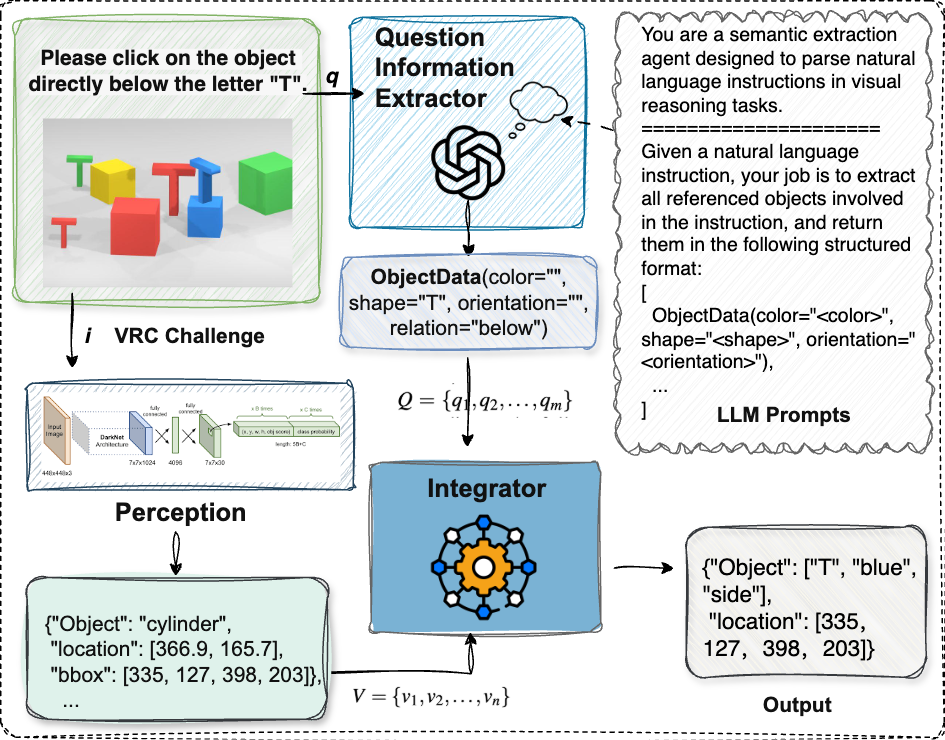}
    \vspace{-15pt}
    \caption{\textbf{Semantic extraction and alignment.} QIE parses the query into slots; the Integrator matches them with detector outputs to yield candidate set $V'$.}
    \label{fig:semantic_alignment}
\end{figure}

\smallskip
\noindent\textbf{Question information extractor.}
QIE reads the natural-language instruction $q$ and turns it into a few short records that name the objects mentioned in the text. Each record has three fields - \emph{shape}, \emph{color}, and \emph{orientation}; when the instruction does not specify a field, that field is left empty and later treated as a wildcard. QIE also notes simple relation phrases (for example, ``\textit{below}'', ``\textit{left of}'') so that the downstream modules know when a spatial step is needed. This flow is shown in Fig.\ref{fig:semantic_alignment}: the instruction is parsed by QIE, then aligned with detections by the Integrator, and, if a relation is present, augmented by the relative-position step. 
We provide two question examples.

($Q1$) For “\textit{Please click on the object directly below the letter ‘T’},” QIE outputs one record for the reference letter with \texttt{shape=T} and empty \texttt{color}/\texttt{orientation}, and it marks the relation “below.” The Integrator will gather all detected \texttt{T}’s, and the RPI module will look underneath each one to recover the target. 

($Q2$) For “\textit{Please click on the number that matches the direction of the red cone},” QIE outputs two records: one for the reference (\texttt{shape=cone}, \texttt{color=red}) and one for the candidate set (\texttt{shape=number} with other fields empty). The Integrator expands ``\textit{number}'' to concrete digit labels, and the Policymaker later applies the ``\textit{same direction}'' check when selecting the final coordinate.

\smallskip
\noindent\textbf{Integrator.}
It serves as a filtering mechanism that bridges the visual output from the multi-object detector and the semantic intent extracted by QIE. Its primary function is to identify, from all detected objects in the image, only those that are semantically relevant to the reasoning task as indicated by the structured QIE output. Integrator acts as a selector, dynamically matching object-level attributes while suppressing irrelevant visual information.

Given the set of detected objects $V = \{v_1, v_2, \dots, v_n\}$ from the detector, where each $v_i$ contains fields such as object, location, and bbox, and a set of structured object queries from QIE: $Q = \{q_1, q_2, \dots, q_m\}$, Integrator performs matching according to the following logic:
\begin{packeditemize}
    \item For each $q_j \in Q$, Integrator selects all $v_i \in V$ such that all non-empty fields in $q_j$ match the corresponding attributes in $v_i$.
    \item If a field in $q_j$ is empty (e.g., \textit{shape = “”}), it is treated as a wildcard and ignored during matching.
    \item For general category descriptions returned by QIE (e.g., ``\textit{letter}'', ``\textit{number}'', ``\textit{3D object}''), Integrator is equipped with predefined type maps to identify relevant visual labels.
\end{packeditemize}

This partial matching scheme enables Integrator to function even when the question is under-specified or refers to abstract categories. 


\begin{algorithm}[!]
\caption{Relative Position Information Extractor}
\label{alg:RPIE}
\small
\begin{algorithmic}[1]
\REQUIRE Instruction $q$; reference set $R$ from Integrator; detections $V$
\ENSURE Augmented set $V_{\text{rel}}$
\STATE $V_{\text{rel}} \leftarrow \emptyset$;\quad $r \leftarrow \texttt{extract\_spatial\_relation}(q)$
\IF{$r = \texttt{None}$} \RETURN $R$ \ENDIF
\FOR{each $r_i \in R$}
  \STATE $p \leftarrow \texttt{center}(r_i.\text{bbox})$;\quad $\hat{p} \leftarrow \texttt{shift}(p, r, \Delta)$
  \FOR{each $v_j \in V$}
    \IF{$\hat{p} \in v_j.\text{bbox}$}
      \STATE add $r_i$ and $v_j$ to $V_{\text{rel}}$;\quad \textbf{break}
    \ENDIF
  \ENDFOR
\ENDFOR
\RETURN $V_{\text{rel}}$
\end{algorithmic}
\end{algorithm}

\subsection{Relative Position Reasoning}
\label{sec:rpie}

Attribute matching alone cannot resolve prompts where the target is defined only by its relation to another object. In such cases, the Integrator retrieves the reference (e.g., the letter ``\texttt{T}'') but the true target (e.g., ``the object directly below the \texttt{T}'') is absent from the candidate set. The Relative Position Information Extractor (RPIE) bridges this gap by translating the linguistic relation into a geometric test over detected bounding boxes.

\smallskip
\noindent\textbf{Relation detection.}
RPIE uses the relation cue flagged by QIE (Fig.\ref{fig:semantic_alignment}). When QIE indicates a spatial predicate such as ``\textit{left of}'', ``\textit{right of}'', ``\textit{above}'', or ``\textit{below}'', RPIE proceeds with geometric projection; otherwise it returns the Integrator output unchanged. If the wording is non-standard and the cue is missing, a lightweight pattern match is applied as a fallback to recover same predicates.

\smallskip
\noindent\textbf{Geometric projection.}
Given a detected relation, let $R$ denote the set of \emph{reference} detections and $V$ the full set of detections. For each reference $r_i\!\in\!R$ with bounding box $b_{r_i}=[x_1,y_1,x_2,y_2]$, RPIE computes its center
\[
p_{r_i}=\Big(\tfrac{x_1+x_2}{2},\,\tfrac{y_1+y_2}{2}\Big).
\]
From the relation token $r$, where the token $r \in \{\textit{left of}, \textit{right of}, \textit{above}, \textit{below}\}$, RPIE derives a direction vector $\vec{d}(r)$ and applies a geometric test on bounding boxes. For example, if $r=\textit{left of}$, the target box must satisfy $x_{\max}(v_t) < x_{\min}(v_r)$, where $v_r$ is the reference object returned by the Integrator. Analogous constraints apply for the other three predicates.
\[
u(r)=
\begin{cases}
(-1,\,0), & r=\textit{left of}\\
(1,\,0),  & r=\textit{right of}\\
(0,\,-1), & r=\textit{above}\\
(0,\,1),  & r=\textit{below}
\end{cases}
\]
and forms the \emph{probe point}
\[
\hat{p}_{r_i}=p_{r_i}+\Delta\,u(r),
\]
where $\Delta>0$ is a fixed offset measured in pixels. RPIE then scans $V$ in a fixed traversal order and selects the first detection $v_j$ whose box contains $\hat{p}_{r_i}$. If such a $v_j$ exists, it is added as an \emph{inferred target} and linked to $r_i$ so that downstream reasoning can see the pair explicitly; if no box contains $\hat{p}_{r_i}$, no target is inferred for that reference.

\smallskip
\noindent\textbf{Coordinate model and tests.}
Boxes are represented by image-space corners $[x_1,y_1,x_2,y_2]$ with the top-left pixel as origin, $x$ increasing to the right and $y$ increasing downward. The operator $\texttt{shift}(p,r,\Delta)$ in Algorithm~\ref{alg:RPIE} is the update $\hat{p}=p+\Delta\,u(r)$. The point-in-box predicate is
\[\scriptsize
\texttt{contains}(\hat{p},[x_1,y_1,x_2,y_2]) \;\triangleq\; (x_1 \le \hat{p}_x \le x_2)\ \wedge\ (y_1 \le \hat{p}_y \le y_2).
\]

If $\hat{p}$ lies outside the image bounds, the check for that reference is skipped. When multiple boxes would contain $\hat{p}$, the first match under the traversal order is taken to keep the procedure stable.

\smallskip
\noindent\textbf{Output.}
RPIE emits an augmented candidate set that contains the original references and any targets inferred by projection, together with reference–target links. This set, along with the instruction, is then passed to the Policymaker, which uses the explicit pairing in a task-aware prompt to select the final coordinate.

\subsection{Prompt-Conditioned Reasoning}
\label{sec:policymaker}

Policymaker is the final stage. It takes the original instruction together with the reasoning cues produced upstream (for example, the spatial predicate flagged by QIE and the reference–target links from RPIE) and constructs a task-aware prompt that directs an LLM to return a single coordinate in image space. The interaction is strictly answer-only: the model is required to output one pair $(x,y)$ in pixel units with the image origin at the top-left.

\begin{figure}[!t]
    \centering
    \includegraphics[width=1.0\linewidth]{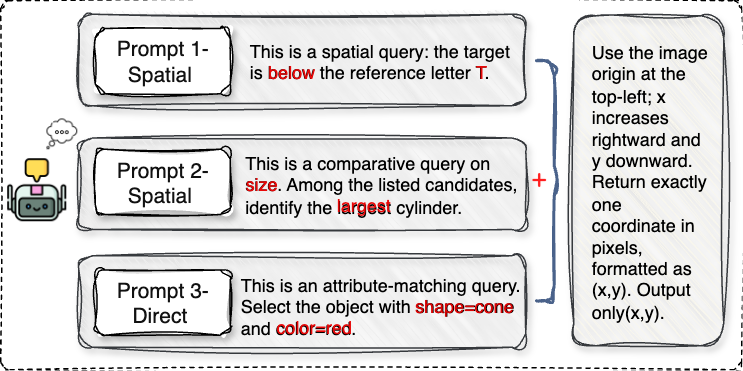}
    \vspace{-0.2in}
    \caption{\textbf{Task-conditioned prompts for Policymaker} (\S\ref{sec:policymaker}). \emph{Spatial} prompt that names the relation and  reference; \emph{Comparative} prompts define an attribute and target class; \emph{Direct-attribute} prompts bind shape and color.}
    \label{fig:prompt_examples}
\end{figure}

\smallskip
\noindent\textbf{Dynamic prompt} (Fig.\ref{fig:prompt_examples}).
Rather than using a fixed template, the prompt is conditioned on the reasoning \textit{type} inferred from the instruction. For \emph{spatial} queries (for example, ``\textit{left of}'', ``\textit{right of}'', ``\textit{above}'', ``\textit{below}''), the prompt names the relation and the reference object(s) and states that the answer must be the coordinate of the object positioned accordingly. For \emph{comparative} queries (for example, ``\textit{the largest sphere}'', ``\textit{the smaller cone}''), the prompt explicitly states the comparison attribute and instructs the model to identify the unique object that satisfies it. For \emph{direct attribute} queries (for example, ``\textit{the red cone}''), the prompt specifies the attribute conjunction and requests the corresponding coordinate. In all cases, the prompt includes an explicit answer schema (``\textit{output exactly one coordinate in the form \texttt{(x,y)} in pixels}''), reminds the model that only the described scene may be used (no unseen objects or assumptions), and clarifies the coordinate convention (integer pixels; $x$ increases rightward, $y$ increases downward). 

\begin{table*}[t]
\centering
\caption{\textbf{Dataset summary} of six real-world VRC platforms.}
\label{tab:dataset_summary}
\scriptsize
\vspace{7pt}

\begin{tabularx}{\textwidth}{c|cc|c|c|X}
\toprule
\multicolumn{1}{c}{\textbf{Dataset}} & \makecell{\textbf{Objects} (per image)} & \multicolumn{1}{c}{\textbf{Object types}} & \multicolumn{1}{c}{\textbf{\quad Prompt variety \quad }} & \multicolumn{1}{c}{\makecell{\textbf{Reasoning}  \textbf{complexity}}} & \multicolumn{1}{c}{\textbf{Typical challenges}} \\
\midrule
\textbf{VTT~\cite{vtt}} & 10–20 & 3D shapes, letters, digits & \cellcolor{green!20}High & \cellcolor{green!20}High & Spatial reasoning, relative position, abstraction \\
\textbf{Xiaodun~\cite{xiaodun}} & 12–14 & Shapes, letters, digits & \cellcolor{yellow!20}Medium & \cellcolor{orange!20}Medium–High & Object stacking, occlusion, attribute + spatial mix \\
\textbf{Geetest~\cite{geetest}} & 7–10  & Regular 3D shapes & \cellcolor{yellow!20}Medium & \cellcolor{yellow!20}Medium & Occlusion, unclear spatial relations \\
\textbf{NetEase~\cite{netease}} & 5–7   & Shapes, letters, digits & \cellcolor{yellow!20}Medium & \cellcolor{yellow!20}Medium & Color/direction filtering, attribute matching \\
\textbf{Dingxiang~\cite{dingxiang}} & 5     & Shapes, letters & \cellcolor{red!20}Low & \cellcolor{orange!15}Low–Medium & Location-only queries, simple relations \\
\textbf{Shumei~\cite{shumei}} & 6     & Regular 3D shapes & \cellcolor{red!20}Low & \cellcolor{red!20}Low & Attribute filtering (smallest + color + shape) \\
\bottomrule
\end{tabularx}
\vspace{0.5em}
\scriptsize
\parbox{\textwidth}{
\textit{Note:} \textbf{Object stacking} refers to CAPTCHA images where multiple objects are vertically stacked, requiring visual disentangling of overlapping or occluded items. 
\textbf{Attribute + spatial mix} indicates questions involving both attribute filtering (e.g., color, shape) and spatial reasoning (e.g., left of, above, below).
}
\vspace{-0.2in}
\end{table*}

\smallskip
\noindent\textbf{LLM inference} (or \textbf{Analyzer}).
Analyzer takes three inputs: the task-conditioned prompt $p$, the image $I$, and the pruned candidate set $V'$. It returns a single pixel coordinate. According to Equation \ref{eq-llm}, we have
\[
f_{\text{LLM}}(p, I, V') \rightarrow (x,y).
\]
In this interface, $p$ specifies the reasoning mode and fixes the answer schema in pixels; $V'$ constrains the decision surface to a small, serialized set of candidates, each with an attribute triple (shape–color–orientation) and a center location. $I$ is provided when visual grounding is needed (e.g., size or pose cues that benefit from the pixels), but Analyzer can often decide from $V'$ alone. The model runs in a single pass; no chain-of-thought is requested.

Within this setup, the system resolves different query types through explicit decision rules.  
For direct attribute queries, it performs exact matching between the specified fields and the attributes in $V'$, and returns the center of the unique match.  
For spatial queries, it relies on the relation and reference–target linkage produced upstream: if a link exists, the system outputs the center of the linked target; if no link is available, Analyzer enforces the stated relation geometrically by comparing object centers along the appropriate axis in $V'$.  
For comparative queries, it computes the requested statistic over the boxes in $V'$: ``\textit{largest}'' is determined by maximizing
\[
\text{area}(b)=(x_2-x_1)\cdot(y_2-y_1),
\]
while ``\textit{leftmost/rightmost}'' is decided by comparing $x$ coordinates of centers, and “topmost/bottommost” by comparing $y$ coordinates.  
In all cases, the chosen candidate’s center $(x,y)$ is emitted as the final answer.

\section{Experiments}
\label{sec-evalua}

\subsection{Settings}
\label{sec:benchmark}
\textbf{Benchmark.} We construct a unified benchmark to ensure fair and reproducible evaluation across systems. The benchmark aggregates public datasets from six commercial VRC providers. Using Selenium, we crawled 12{,}000 samples from Tencent’s official VTT service~\cite{vtt} and incorporated five additional datasets from the open-source VRC benchmark~\cite{xu2023vrc}: Xiaodun (12{,}000 samples), Geetest (12{,}000), NetEase (12{,}000), DingXiang (12{,}000), and Shumei (3{,}000). Together, these datasets span diverse visual styles, question formats, and difficulty levels, providing broad coverage of real-world VRC deployments (cf.~Table~\ref{tab:dataset_summary}). We publicly release the complete benchmark, including raw data and annotations\footnote{\url{https://zenodo.org/records/17971722}} to facilitate follow-up studies. Appendix~\ref{sec:data_sources} details data formats and challenge characteristics of each provider.

\textit{Normalization.}
We restructure all datasets into a unified object-centric vision–language format. Raw inputs are converted into a standardized detector-compatible representation with per-object bounding boxes annotated by color, shape, and orientation. For datasets lacking box annotations, objects are manually labeled using a custom interface to ensure cross-source consistency.
For evaluation, we construct a ``one-answer'' ground-truth set by uniformly sampling 1{,}000 images per provider and annotating a single correct click region per image. A prediction is considered correct if the output coordinate falls within this region. All models, baselines, and ablations are evaluated under this data format and scoring protocol.

\smallskip
\noindent\textbf{Baselines}. We compare \textsc{ViPer} against three representative VRC solvers capturing dominant attack paradigms: perception-driven, graph-based reasoning, and LLM-driven pipelines. For consistency, all baselines were re-implemented on our benchmark to ensure fairness.

\begin{packeditemize}
    \item\textbf{Vision-based solvers.}  
The Holistic model~\cite{gao2021research} pioneered modular attacks by combining detection with rule-based logic, reaching up to 88\% success on platforms such as Tencent VTT. We re-implemented this pipeline and its extended version~\cite{wang2023extended}. In parallel, VRC-GraphNet~\cite{xu2023vrc} encodes detections as visual–language graphs, where nodes represent objects and edges capture spatial or semantic relations. A graph neural network is then trained to select the target node. We adapted this approach by normalizing detections and retraining the GNN on our benchmark.

    \item\textbf{LLM-driven solver.}  
Oedipus~\cite{deng2024Oedipus} represents reasoning-centric attacks, translating VRC queries into a domain-specific language and applying chain-of-thought prompting with an LLM. We reimplemented the full pipeline and evaluated it, aligning the backbone with \textsc{ViPer} to isolate framework-level differences.  
\end{packeditemize}

\subsection{Method Variants For Ablation}
\begin{figure}[!]
    \centering
    \subfigure[\textbf{LLM-only (R1)}: the instruction and image are passed directly to the multimodal LLM; no structured perception or alignment is used.]{
        \label{fig:r1_schematic}
        \includegraphics[width=1.0\linewidth]{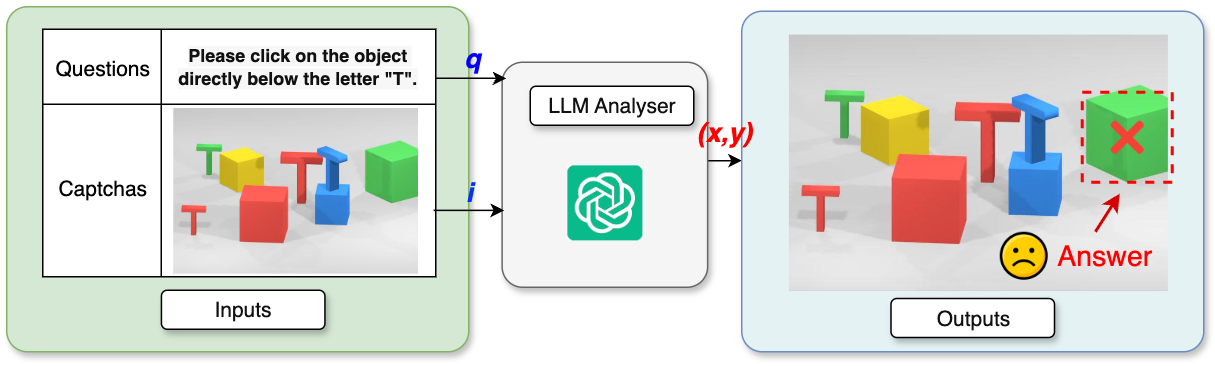}
    }
    \vspace{0.6em}
    \subfigure[\textbf{Detector$\rightarrow$Prompt (R2)}: objects are detected and outputs (labels, boxes, centers) are injected into prompts, without QIE/Integrator/RPIE.]{
        \label{fig:r2_schematic}
        \includegraphics[width=1.0\linewidth]{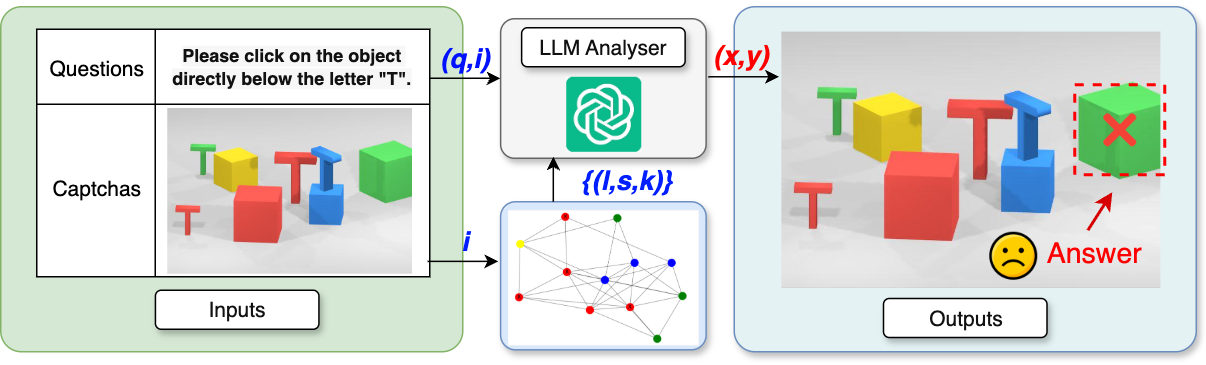}
    }
    \vspace{-15pt}
    \caption{\textbf{Schematic of ablation variants.} R1 relies solely on native multimodal vision and zero-shot reasoning. R2 augments the prompt with detector outputs but omits semantic filtering and relative-position inference.}
    \label{fig:r1r2_schematic}
\end{figure}

We define two reduced solvers (Fig.\ref{fig:r1r2_schematic}) that differ from \textsc{ViPer} by removing specific components, while keeping all other settings identical.

\smallskip
\noindent\textbf{R1: LLM-only.} The instruction $q$ and image $I$ are given directly to a multimodal LLM without any structured perception or alignment. The prompt is minimal and asks for a single image-space coordinate $(x,y)$. There are no detector outputs and no QIE/Integrator/RPIE steps. This variant relies entirely on the model’s native vision to parse the scene and resolve the query, which makes it prone to errors with small or repeated objects, fine-grained attributes such as color or pose, and spatial relations expressed only in text.

\smallskip
\noindent\textbf{R2: Detector\,$\rightarrow$\,Prompt.} 
The detector lists all objects in $I$ with compound labels, boxes, and centers. This unfiltered list is appended to the prompt instruction. No QIE parsing, Integrator pruning, or RPIE inference is applied. LLM is asked to return one $(x,y)$. While this adds explicit object candidates, the prompt often includes many irrelevant or near-duplicate entries and lacks explicit handling of relational cues, which can lead to confusion when attributes partially match or when the target is defined only by position with respect to a reference. 

\subsection{\textsc{ViPer} Evaluation}
\label{sec:evaluation}
\smallskip
\noindent\textbf{Configurations.}  Our experiments are conducted on a consistent, public benchmark spanning six commercial VRC platforms (\S\ref{sec:benchmark}). The environment includes a 15-core Intel Xeon Platinum 8474C CPU and an NVIDIA RTX 4090D GPU (24GB) for model training, and an AMD Ryzen 9 7845HX CPU with RTX 4060 (8GB) for inference. We use PyTorch v2.5.1 with CUDA 12.1, and all LLM calls are made via public APIs. 

We evaluate \textsc{ViPer} under four dimensions: 

\begin{packeditemize}
    \item[\ding{172}] \textit{Component ablation.} To isolate design contributions, we compare three progressive variants: R1 (base viewer only), R2 (adds relation inference), and R3 (full \textsc{ViPer}). Accuracy is measured across all datasets using consistent image-question inputs.

    \item[\ding{173}] \textit{Baseline comparison.}  
    We benchmark \textsc{ViPer} (GPT-4o backend) against three prior CAPTCHA solvers: Holistic model~\cite{gao2021research, wang2023extended}, VRC-GraphNet~\cite{xu2023vrc}, and Oedipus~\cite{deng2024Oedipus}, on the same benchmark. In addition, we report a human baseline obtained from ten participants solving randomly sampled challenges from the identical benchmark corpus. This baseline provides an empirical upper bound on achievable accuracy and a point of reference for model latency (cf. \S\ref{app:human}).
    
    \item[\ding{174}] \textit{LLM backend swap.} To evaluate the robustness and portability of \textsc{ViPer}’s reasoning module, we test it with four state-of-the-art (SOTA) LLM backends: GPT-4o, DeepSeek-Chat, Kimi-latest, and Grok-2. Each model receives the same dynamic prompt and Viewer outputs, with no chain-of-thought enabled. 
    
    \item[\ding{175}] \textit{Response time.}  
    We measure average end-to-end inference time (per image-question pair) across all LLM backends to evaluate latency tradeoffs.
\end{packeditemize}

\smallskip
\noindent\textbf{Experimental results.}  All experimental results across the three testing protocols are summarized in Table~\ref{tab:captcha_accuracy}.

\begin{table}[!htbp]
\centering
\scriptsize 
\renewcommand{\arraystretch}{1.1}
\caption{\textbf{Accuracy (\%) of various benchmarks across six VRC platforms} (\textcolor{green!20}{\rule{6pt}{6pt}} = overall best; 
\textcolor{orange!20}{\rule{6pt}{6pt}} = overall second; \textcolor{red!20}{\rule{6pt}{6pt}} = overall worst; -- = no data)}
\label{tab:captcha_accuracy}
\vspace{7pt}
\begin{tabularx}{\columnwidth}{c|>{\columncolor{kimicolor}}c|YYYYYY}
\toprule
\textbf{Benchmark} & \multicolumn{1}{c}{\textbf{Backend}} & \textbf{VTT} & \textbf{Geetest} & \textbf{NEase} & \textbf{DingX} & \textbf{Shumei} & \textbf{Xiaod} \\
\midrule
Holistic~\cite{gao2021research,wang2023extended} & -- & 88.12 & 85.74 & 86.37 & 88.52 & 95.84 & 79.28 \\
GraphNet~\cite{xu2023vrc} & -- & \cellcolor{orange!20}89.47 & 76.92 & 72.18 & 85.34 & \cellcolor{green!20}99.87 & 89.65 \\
\midrule
\multirow{4}{*}{Oedipus~\cite{deng2024Oedipus}} 
& gpt4o    & 70.23 & 61.47 & 66.82 & 73.16 & 58.44 & 64.08 \\
& deepseek & \cellcolor{red!20}63.72 & \cellcolor{red!20}55.93 & \cellcolor{red!20}60.37 & \cellcolor{red!20}68.14 & \cellcolor{red!20}57.26 & \cellcolor{red!20}59.53 \\
& kimi     & 72.48 & 65.27 & 69.15 & 74.63 & 62.83 & 70.38 \\
& grok-2   & 66.42 & 59.87 & 62.74 & 70.94 & 60.18 & 63.27 \\
\midrule
Human & -- & 87.60 & \cellcolor{green!20}90.70 & \cellcolor{green!20}95.12 & \cellcolor{green!20}95.40 & 93.20 & \cellcolor{orange!20}91.80 \\
\midrule
\multirow{4}{*}{Ablation-R1} 
& gpt4o    & 31.36 & 26.80 & 16.41 & 24.87 & 7.07 & 10.99 \\
& deepseek & 28.42 & 24.36 & 18.47 & 23.12 & 8.63 & 11.82 \\
& kimi     & 32.74 & 27.92 & 19.88 & 24.97 & 9.12 & 12.96 \\
& grok-2   & 29.18 & 27.44 & 17.32 & 26.38 & 6.77 & 10.46 \\
\midrule
\multirow{4}{*}{Ablation-R2} 
& gpt4o    & 75.79 & 61.75 & 66.86 & 73.98 & 77.08 & 62.31 \\
& deepseek & 69.75 & 58.24 & 63.12 & 70.46 & 77.84 & 60.24 \\
& kimi     & 67.85 & 55.36 & 61.42 & 68.24 & 70.42 & 58.33 \\
& grok-2   & 72.21 & 63.15 & 64.53 & 74.82 & 72.24 & 59.67 \\
\midrule
\multirow{4}{*}{\textbf{R3 (\textsc{Viper})}} 
& gpt4o    & \cellcolor{green!20}97.33 & \cellcolor{orange!20}86.45 & \cellcolor{orange!20}92.61 & \cellcolor{orange!20}91.48 & \cellcolor{orange!20}98.45 & \cellcolor{green!20}92.59 \\
& deepseek & 92.96 & 83.25 & 93.00 & 87.88 & 97.50 & 87.76 \\
& kimi     & 88.00 & 85.20 & 91.92 & 89.20 & 95.50 & 87.86 \\
& grok-2   & 92.50 & 81.91 & 89.00 & 86.00 & 96.50 & 92.00 \\
\bottomrule
\end{tabularx}
\end{table}

\begin{figure*}[t]
    \centering
    \subfigure[GPT-4o]{
        \includegraphics[width=0.48\linewidth]{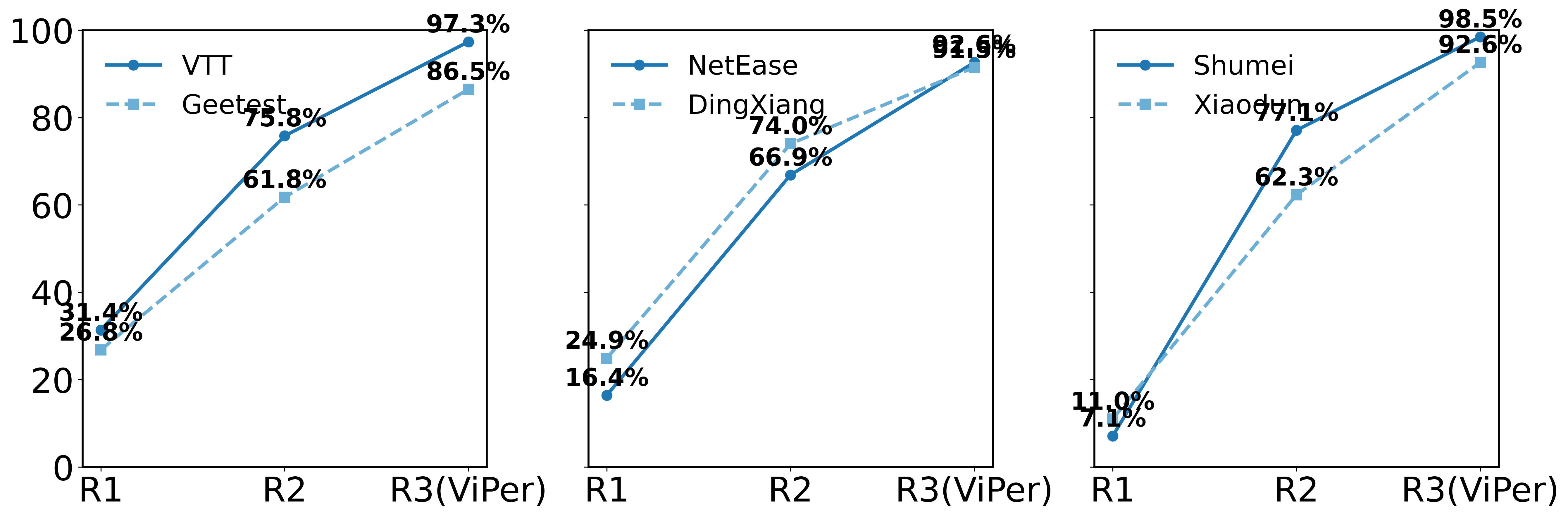}
    }
    \hfill
    \subfigure[DeepSeek-Chat]{
        \includegraphics[width=0.48\linewidth]{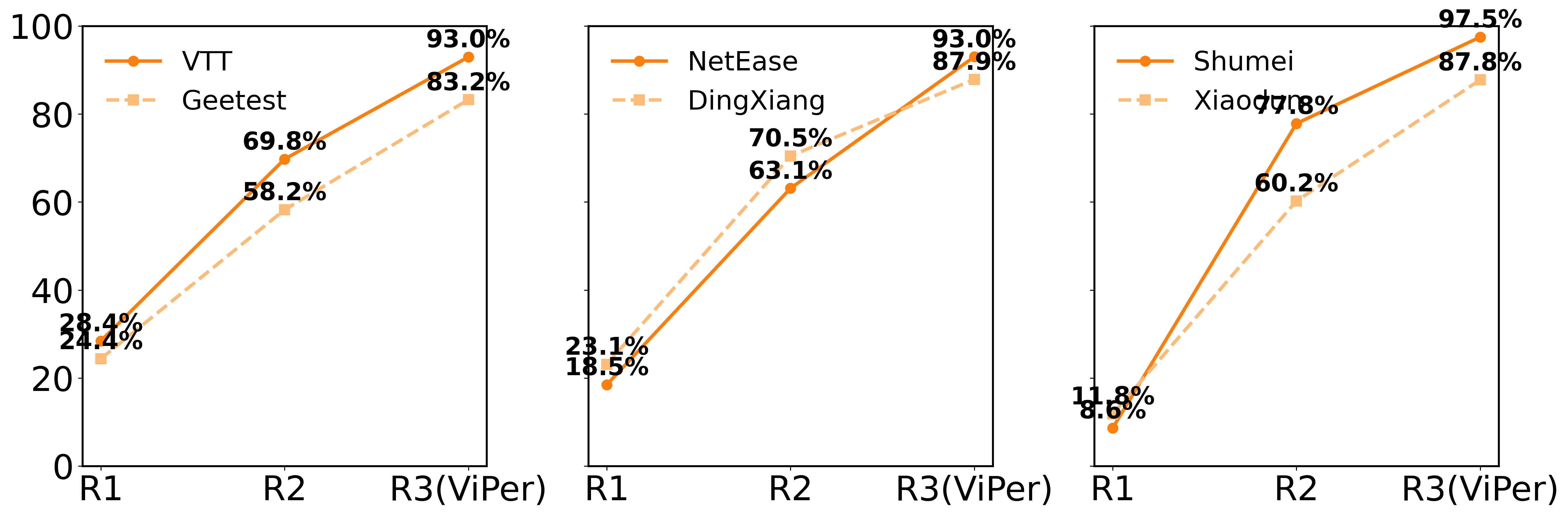}
    }

    \vspace{-0.1cm}

    \subfigure[Kimi-latest]{
        \includegraphics[width=0.48\linewidth]{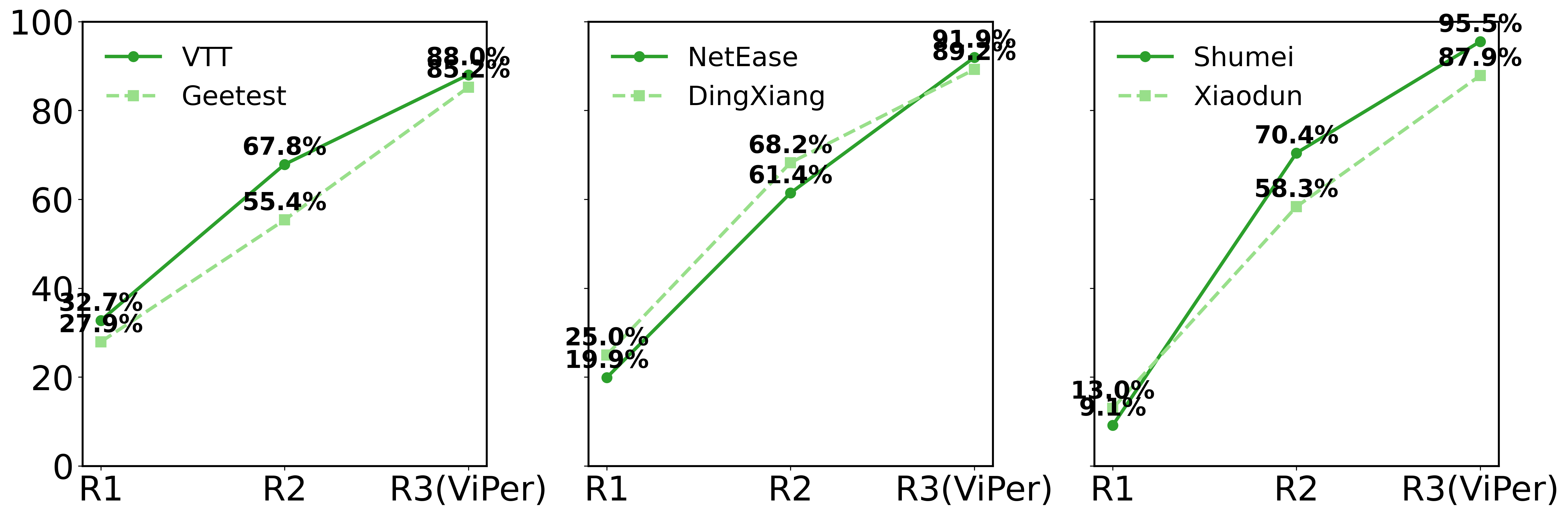}
    }
    \hfill
    \subfigure[Grok-2]{
        \includegraphics[width=0.48\linewidth]{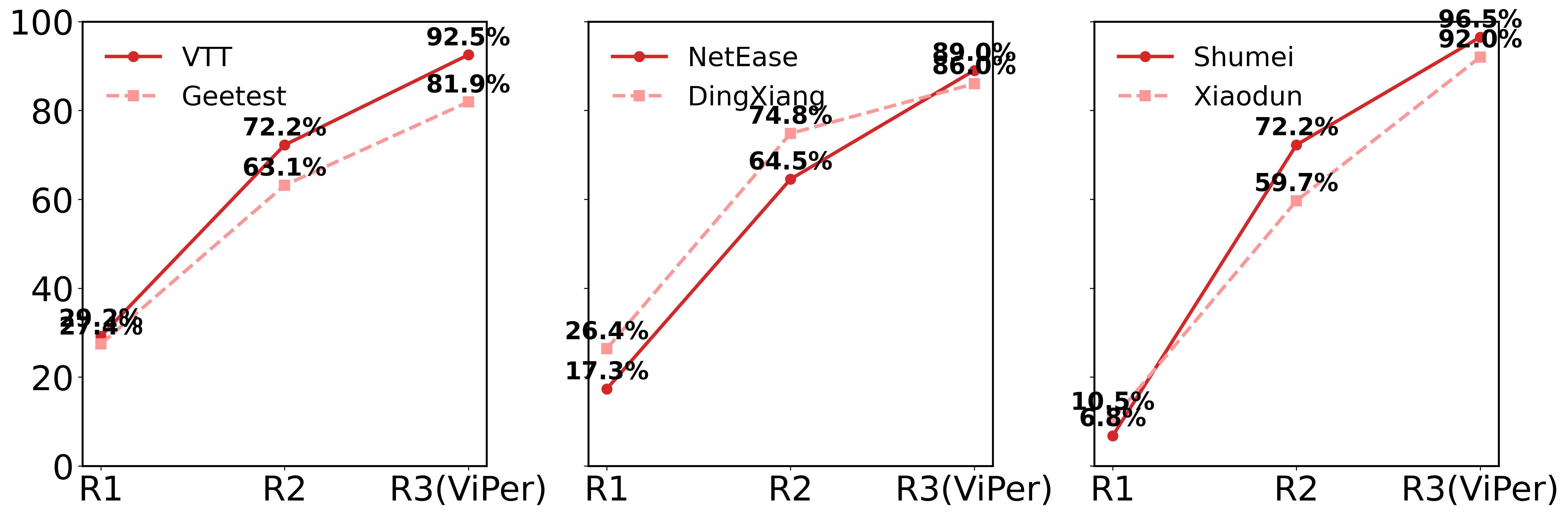}
    }

    \vspace{-0.15in}
    \caption{\textbf{Component-wise ablation of the \textsc{ViPer} framework across four LLM backends.} 
    Each panel reports accuracy for R1 (LLM-only reasoning), R2 (with structured perception), and R3 (full \textsc{ViPer}).}
    \label{fig:comparison_stage}
        \vspace{-0.15in}
\end{figure*}

\noindent\underline{\textit{\ding{172} Component ablation}} (Fig.\ref{fig:comparison_stage}).  
Our ablation analysis demonstrates that each incremental component in \textsc{ViPer} produces consistent accuracy improvements across all six VRC platforms. With GPT-4o as the backend, the baseline configuration (R1), which consumes only raw visual input, yields poor accuracy, ranging from $7.1\%$ on Shumei to $31.4\%$ on VTT. Incorporating structured perception via detection in R2 substantially boosts performance, with all platforms surpassing $60\%$ accuracy and Shumei reaching $77.1\%$. The full model (R3) delivers a further leap, achieving $86.5$--$98.5\%$ across platforms. This shows the necessity of combining detection, spatial reasoning, and hierarchical prompting, particularly for resolving relational queries in cluttered visual scenes.   

The effectiveness of the ablation scheme generalizes to other backends. DeepSeek follows the same upward trajectory, rising from $8.6$--$28.4\%$ (R1) to $58.2$--$77.8\%$ (R2), and finally reaching $83.3$--$97.5\%$ (R3). Kimi begins with a slightly stronger baseline (up to $32.7\%$ on VTT) but converges to a lower ceiling of $85.2$--$95.5\%$ in R3, suggesting weaker compatibility with structured prompts. Grok-2 also demonstrates steady gains, advancing from $6.8$--$29.2\%$ (R1) to $59.7$--$74.8\%$ (R2), and ultimately reaching $81.9$--$96.5\%$ in R3. Although backend-specific variations exist, such as Kimi’s weaker plateau and Grok’s strong showing on Geetest and DingXiang, the consistent monotonic improvement from R1 through R3 across all backends confirms that our staged integration strategy is both effective and robust.  

\textcolor{teal}{\textbf{$\circ$ Analysis.}} 
Beyond aggregate accuracy, the ablation results reveal clear failure patterns across stages. The sharp jump from R1 to R2 indicates that errors in R1 are dominated by perceptual failures: LLMs operating directly on raw images frequently misidentify small, repeated, or visually similar objects, leading to incorrect reference selection. While R2 mitigates most of these errors by grounding reasoning in explicit object candidates, residual failures persist when multiple objects partially satisfy the query attributes, particularly for position-only relations. The additional gains from R2 to R3 suggest that these remaining errors are primarily reasoning-related rather than perceptual. Hierarchical prompting and spatial inference help disambiguate near-miss candidates by explicitly encoding relational constraints.

\begin{figure*}[!htbp]
    \centering
    \subfigure[VTT]{
        \includegraphics[width=0.25\textwidth]{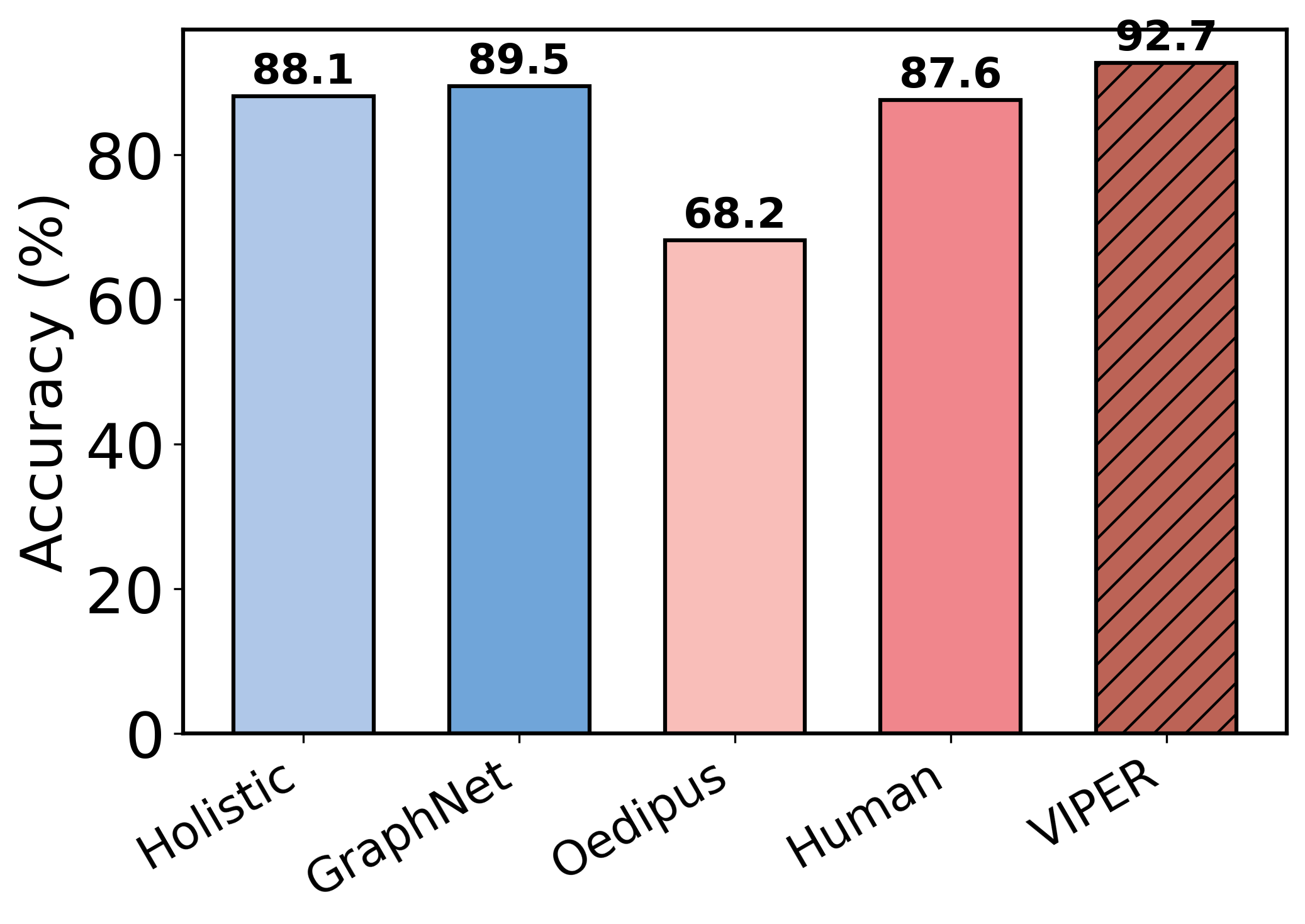}
    }
    \hspace{0.3cm}
    \subfigure[Geetest]{
        \includegraphics[width=0.25\textwidth]{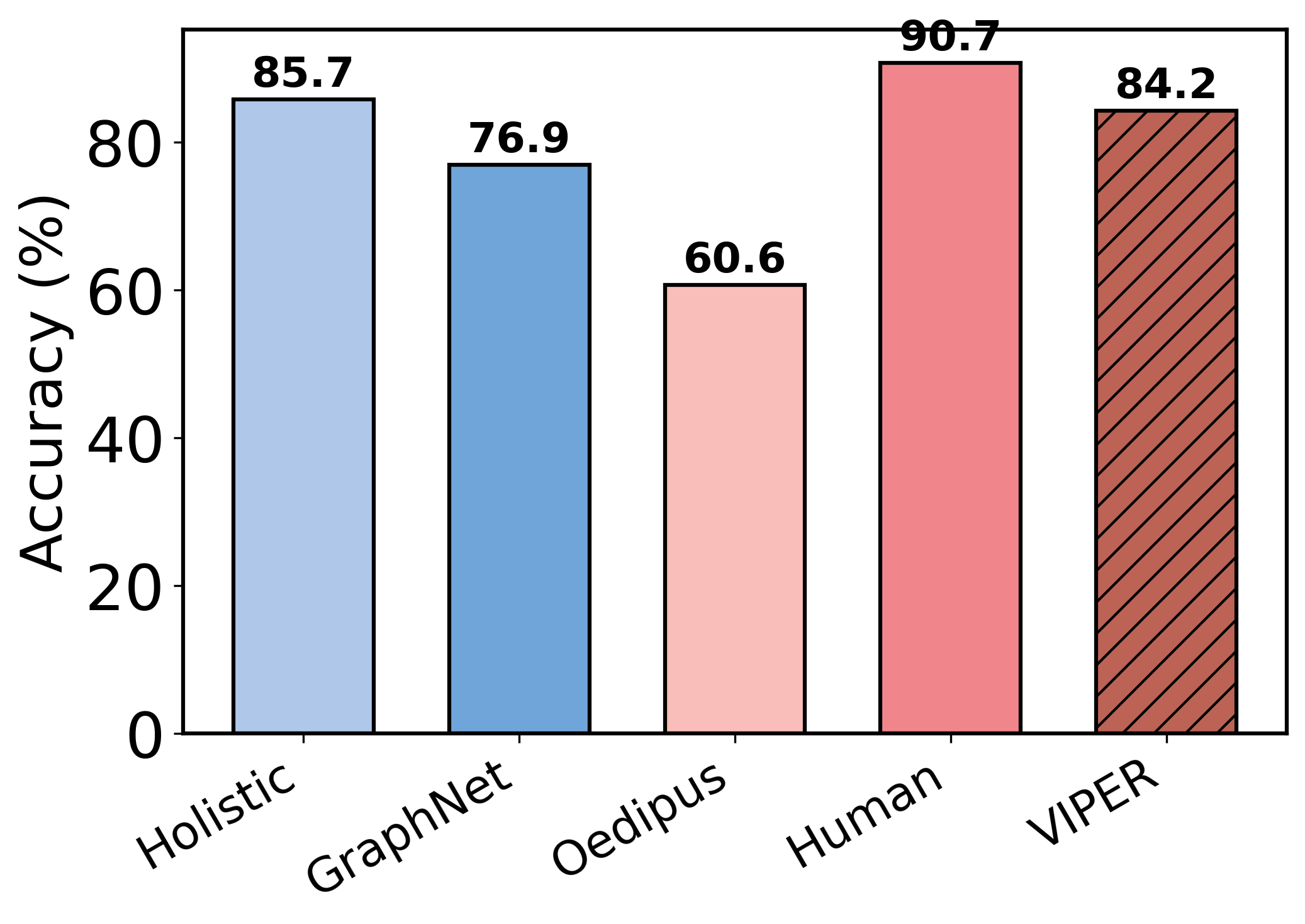}
    }
    \hspace{0.3cm}
    \subfigure[NetEase]{
        \includegraphics[width=0.25\textwidth]{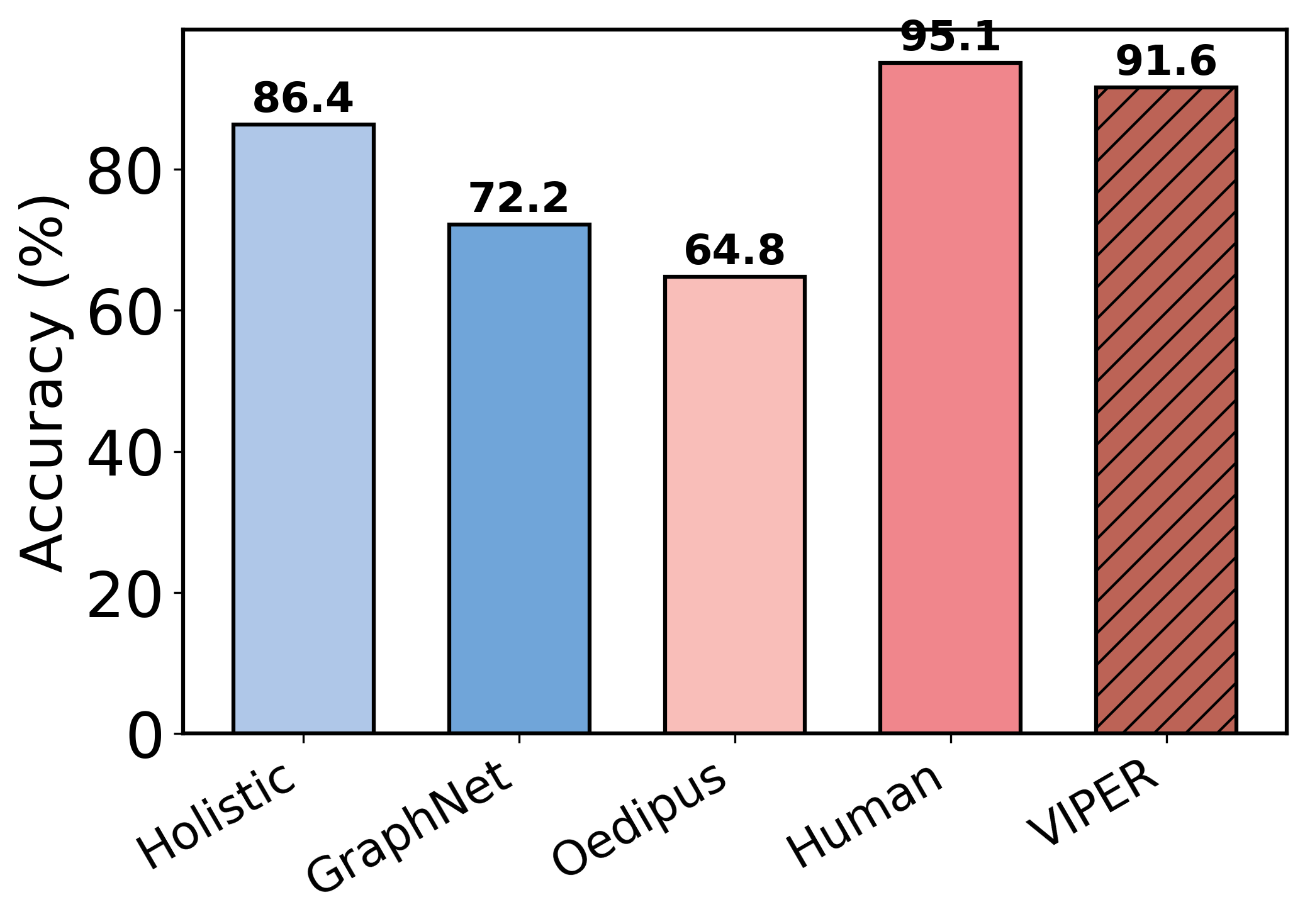}
    }
    \subfigure[DingXiang]{
        \includegraphics[width=0.25\textwidth]{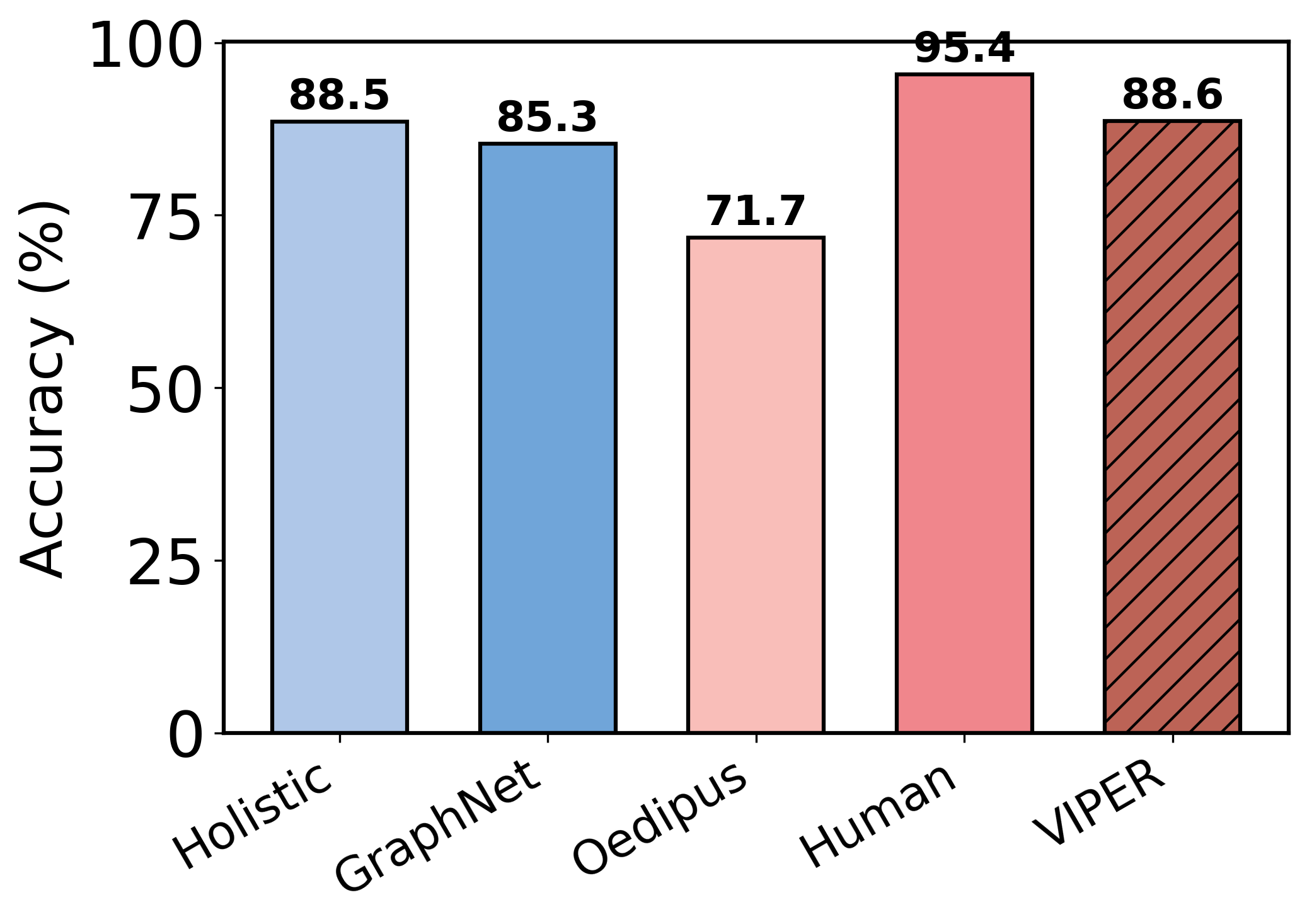}
    }
    \hspace{0.3cm}
    \subfigure[Shumei]{
        \includegraphics[width=0.25\textwidth]{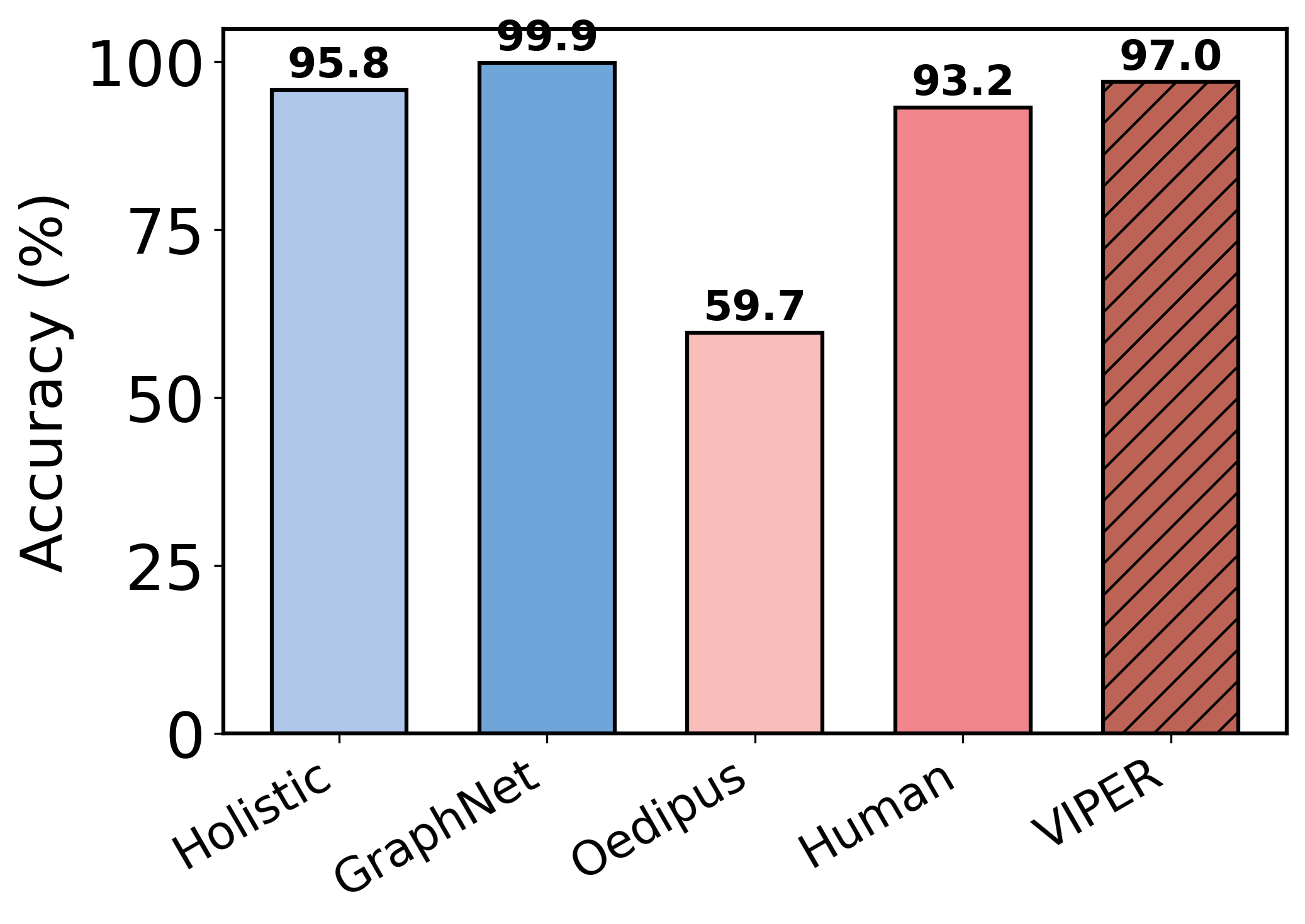}
    }
    \hspace{0.3cm}
    \subfigure[Xiaodun]{
        \includegraphics[width=0.25\textwidth]{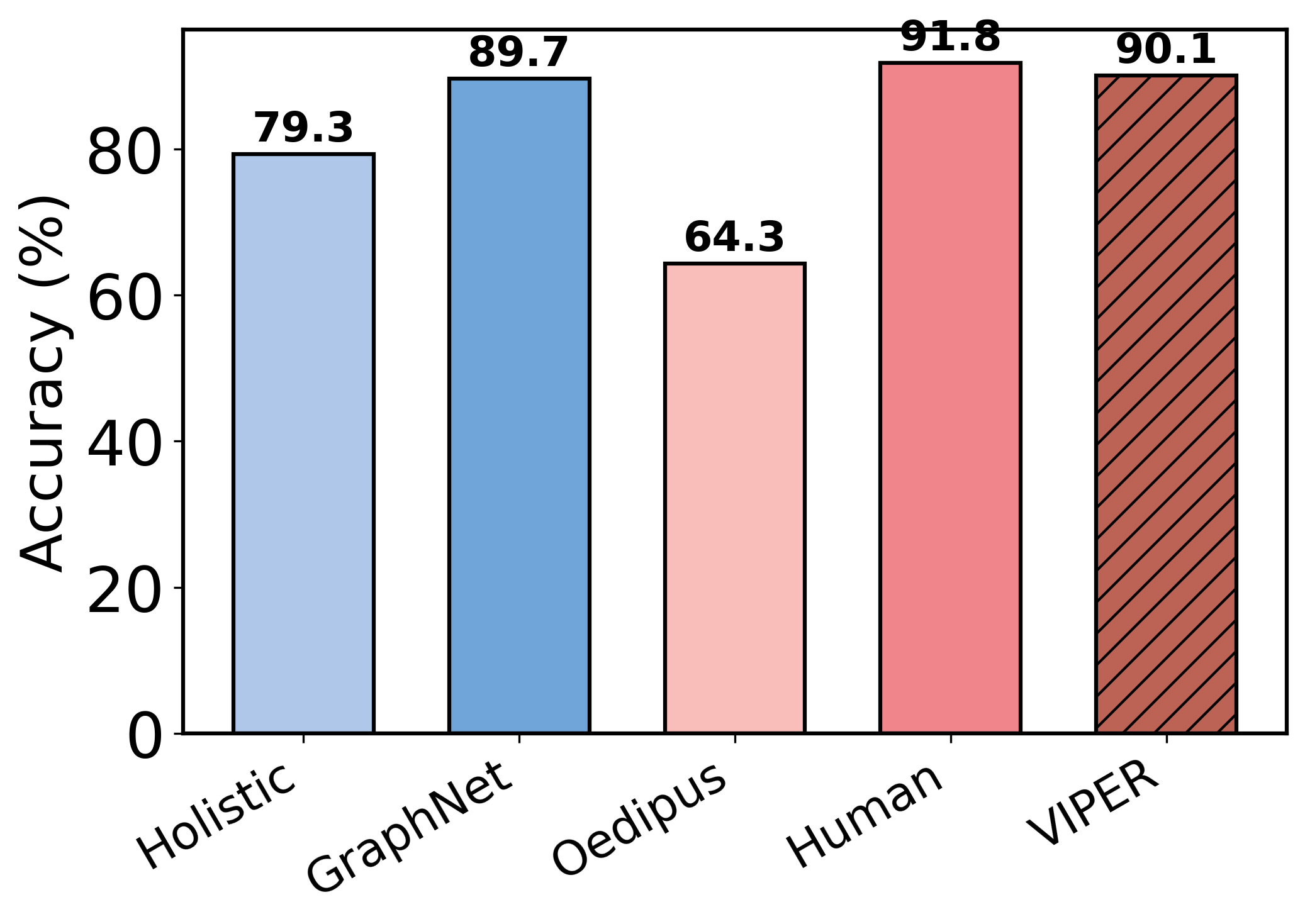}
    } 
    \caption{\textbf{Comparison with benchmarks on six VRC platforms.} 
    Each subfigure reports average accuracy across different methods, highlighting VIPER with hatched bars.}
    \label{fig:compare_benchmarks}
\end{figure*}

\smallskip
\noindent\underline{\textit{\ding{173} Compare with benchmarks}} (Fig.\ref{fig:compare_benchmarks}).  
\textsc{ViPer} establishes new SOTA performance across all six VRC benchmarks, consistently outperforming prior methods and closely matching human-level accuracy.  

On the VTT dataset, the most reasoning-intensive benchmark, \textsc{ViPer} (GPT-4o backend) reaches 97.33\% accuracy, slightly exceeding the human baseline (87.60\%) by 10 percentage points and outperforming Oedipus (70.23\%), GraphNet (89.47\%), and Holistic (88.12\%). This marks an absolute gain of more than 27 points over the strongest prior LLM-based solver.  

On Geetest, \textsc{ViPer} reaches 86.45\%, close to the human baseline (90.70\%) and well above Oedipus (61.47\%) and GraphNet (76.92\%). On NetEase, which emphasizes orientation and color matching, it records 92.61\%—just below humans (95.12\%) but far ahead of Oedipus (66.82\%) and GraphNet (72.18\%). On DingXiang, with simpler spatial relations, \textsc{ViPer} achieves 91.48\%, narrowing the gap to humans (95.40\%) and outperforming Oedipus (73.16\%) by over 18 points.

On Shumei, characterized by low object diversity and repetitive prompts, \textsc{ViPer} reaches 98.45\%, slightly above the Holistic method (95.84\%) and vastly superior to Oedipus (58.44\%). Finally, on Xiaodun, \textsc{ViPer} achieves 92.59\%, outperforming Oedipus (64.08\%) and aligning closely with human performance (91.80\%).  

\textcolor{teal}{\textbf{$\circ$ Analysis.}}
The benchmark results show that \textsc{ViPer}’s improvements are systematic rather than dataset-specific. The largest gains appear on reasoning-intensive benchmarks (e.g., VTT and NetEase), where multi-object relations and compositional constraints dominate, suggesting that structured perception alone is insufficient without reasoning. The remaining gap to human performance on some platforms (e.g., Geetest and DingXiang) is largely due to ambiguous or under-specified queries, where multiple objects satisfy the constraints. In such cases, humans often rely on contextual bias, whereas \textsc{ViPer} adheres strictly to formal query semantics. The errors therefore reflect inherent ambiguity in challenge design rather than failures of perception or reasoning.

\begin{figure}[!t]
    \centering
    \includegraphics[width=0.85\linewidth]{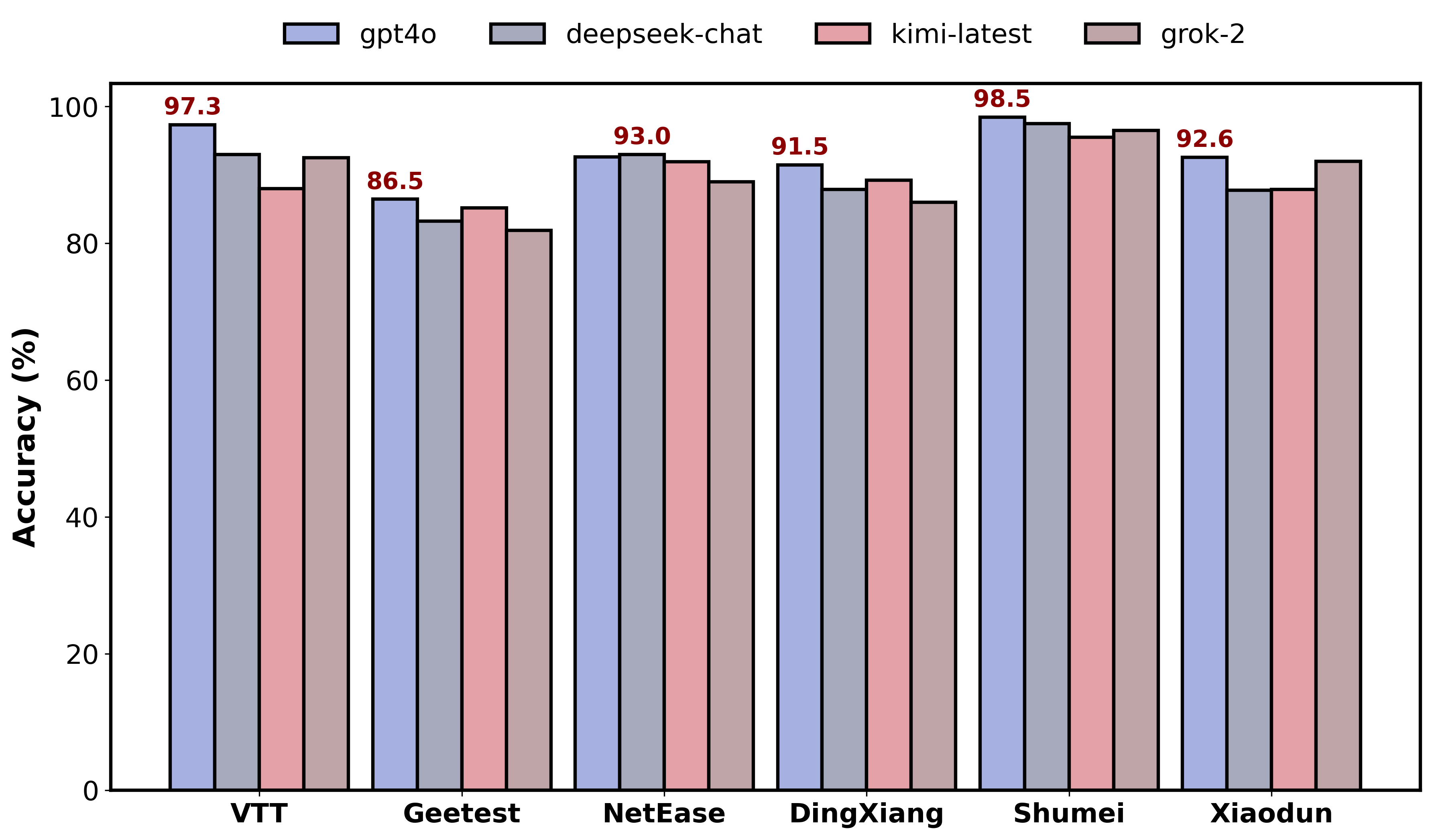}
    \vspace{-0.1in}
    \caption{\textbf{Accuracy comparison of four LLM backbones.} All solvers share the same visual perception and reasoning pipeline, with only the LLM backend varied.}
    \label{fig:llm_comparison}
\end{figure}

\smallskip
\noindent\underline{\textit{\ding{174} LLM backbone comparison}} (Fig.\ref{fig:llm_comparison}).
\textsc{ViPer} with GPT-4o delivers the most consistent and top-performing results. It achieves leading accuracy on VTT (97.33\%), Geetest (86.45\%), Dingxiang (91.48\%), Shumei (98.45\%), and Xiaodun (92.59\%), showing both robustness on high-complexity reasoning tasks and reliability on simpler spatial challenges. \textsc{ViPer} with DeepSeek shows competitive performance, surpassing others on NetEase (93.00\%) and maintaining strong scores on VTT (92.96\%) and Shumei (97.50\%). However, its accuracy fluctuates more across benchmarks, with lower values on Geetest (83.25\%) and Dingxiang (87.88\%).  

\textsc{ViPer} with Kimi exhibits balanced but less dominant results: it performs reliably across datasets, achieving 91.92\% on NetEase and 85.20\% on Geetest, but falls short of leading on any single platform. Its relative underperformance on VTT (88.00\%) and Xiaodun (87.86\%) highlights its limited strength in reasoning-intensive or object-dense settings. Finally, \textsc{ViPer} with Grok-2 attains strong performance on Xiaodun (92.00\%) and Shumei (96.50\%), rivaling the top-performing models in these cases. Yet, its results are uneven, with noticeable degradation on Dingxiang (86.00\%) and Geetest (81.91\%), marking the weakest scores in the cohort.  

\textcolor{teal}{\textbf{$\circ$ Analysis.}} 
The backbone comparison shows that \textsc{ViPer} benefits from stronger LLMs, but does not rely on them exclusively. GPT-4o achieves the best overall performance, particularly on reasoning-intensive benchmarks such as VTT and Geetest, indicating stronger robustness to compositional queries. DeepSeek performs well on attribute-focused datasets (e.g., NetEase) but shows higher variance across platforms, suggesting sensitivity to spatial ambiguity. Kimi and Grok-2 exhibit more uneven performance, with lower performance ceilings or sharp drops on complex layouts, pointing to weaker compatibility with structured perception outputs. These suggest backend-specific failure modes rather than a single dominant bottleneck.

\begin{figure}[!t]
    \centering
    \includegraphics[width=0.8\linewidth]{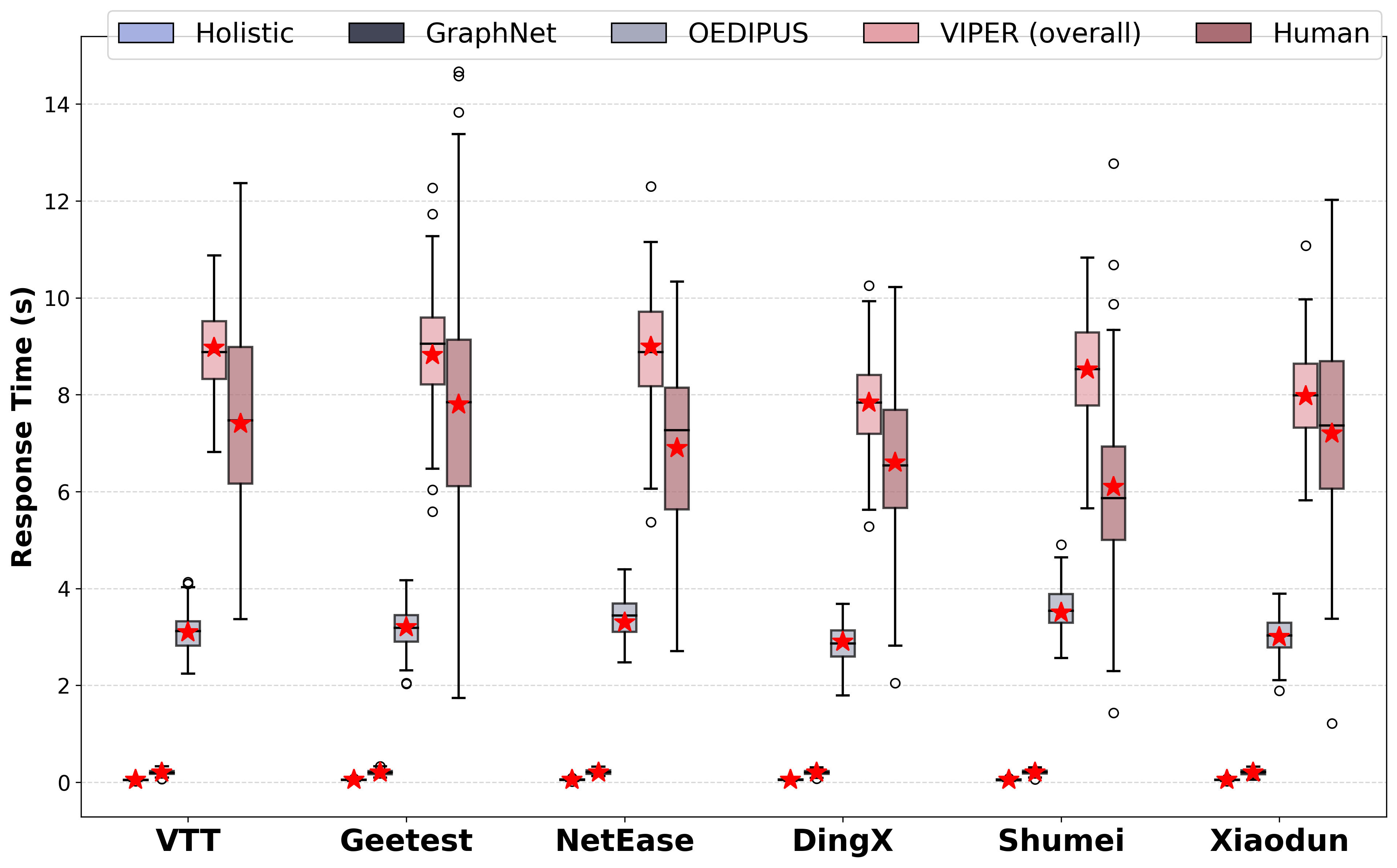}
    \vspace{-0.1in}
    \caption{\textbf{Response time comparison} of mean response times across six VRC platforms by over five solvers.}
    \label{fig:llm_comparison_time}
    \vspace{-0.1in}
\end{figure}

\smallskip
\noindent\underline{\textit{\ding{175} Response time comparison}} (Fig.\ref{fig:llm_comparison_time}).
We further compare the mean response times of \textsc{ViPer} and prior VRC solvers to assess the computational overhead. Existing approaches such as Holistic and GraphNet exhibit extremely low latency, with average completion times below $0.2$ seconds. However, these methods achieve this speed at the cost of accuracy, as they rely on shallow heuristics or graph-based matching rather than full reasoning pipelines. Oedipus, in contrast, demonstrates longer response times (3 seconds on average) while integrating more advanced perceptual modules, but still falls short in solving high-complexity challenges.  

\textsc{ViPer} incurs higher latency due to its multi-stage reasoning pipeline powered by LLMs. Depending on the backbone, response times range from $5$--$11$ seconds (Table~\ref{tab:response_time_summary}), with GPT-4o providing the fastest responses (average 5--6 seconds) and Kimi and DeepSeek producing slower executions (9$-$11 seconds). Grok-2 offers a middle ground, averaging $7$--$9$ seconds. Notably, the overall latency of \textsc{ViPer} (8.3 seconds across platforms) remains comparable to that of the human baseline (7.0 seconds), highlighting that the additional computational cost is of the same order of magnitude as human problem-solving.

\textcolor{teal}{\textbf{$\circ$ Analysis.}}
The response-time comparison highlights an inherent trade-off between latency and reasoning capability. Lightweight solvers such as Holistic and GraphNet achieve sub-second responses by relying on shallow heuristics, which explains their poor performance on reasoning-intensive VRCs. In contrast, \textsc{ViPer} incurs higher latency due to perception and LLM inference stages, but this overhead enables substantially higher accuracy. Notably, \textsc{ViPer}'s end-to-end latency remains comparable to human solving time, suggesting that its computational cost reflects the intrinsic complexity of the task rather than inefficiency. 

\subsection{Multi-object Detector Evaluation}

\noindent\textbf{Training and architecture} (Table~\ref{tab:vrc_labels}).
We trained a unified multi-object detector on a curated pool of 1,200 annotated VRC images, equally sampled from six providers: VTT, DingXiang, Geetest, NetEase, Shumei, and Xiaodun (200 images each). The dataset was partitioned into 800 training, 200 validation, and 200 held-out test samples to ensure balanced evaluation. All images were resized to $640 \times 640$ pixels prior to training. 

The detector adopts a modern single-stage, anchor-free YOLO-style architecture (YOLOv11~\cite{yolov11}), consisting of a convolutional backbone for multi-scale feature extraction, a feature aggregation neck, and a decoupled detection head for classification and bounding-box regression. Each detected object is assigned a compound label encoding shape, color, and orientation. Training was performed on a single NVIDIA RTX 4090 GPU using the AdamW optimizer with a batch size of 32 for 3000 epochs. 

\smallskip
\noindent\textbf{Evaluation} (Table~\ref{tab:yolo_metrics}). 
To assess the performance of our multi-object detector, we report both overall COCO-style metrics and stratified results by object scale and attribute. 
The detector achieves a precision of 0.998 and recall of 0.956, yielding an mAP@50 of 0.984 and mAP@50:95 of 0.919. 
Stratified analysis reveals stable performance across scales, with AP$_s$ = 0.912 for small objects, AP$_m$ = 0.944 for medium, and AP$_l$ = 0.963 for large objects, indicating strong robustness to variations in target size. 
Attribute-level indicators such as AR@100 and TP/FP per image further highlight the detector’s ability to capture fine-grained features with minimal false positives. 
 
In addition, per-platform evaluations (Part C) demonstrate competitive cross-domain generalization: for instance, mAP@50:95 reaches 0.965 on VTT and 0.951 on DingX, while remaining 0.907 on Geetest and 0.857 on Xiaodun. 
The detector also achieves near-perfect mAP@50 across all platforms ($\geq0.95$), with a mean $\pm$ std of $0.932 \pm 0.009$ across six validation platforms. 
Runtime profiling (Part D) further confirms the efficiency of the model, with 25.6M parameters and 72.4 GFLOPs at input size 640, yielding 7.8 ms latency per image on GPU (128 images/s) and a lightweight 0.9 ms NMS overhead. 
Additionally, Fig.\ref{fig:train_results} and Fig.\ref{fig:yolo_accuracy} in Appendix~\ref{app:detector} illustrate representative qualitative performance.

\smallskip
\noindent\textbf{Ontology scope and generalization discussion.}
The multi-object detector in \textsc{ViPer} operates over a fixed ontology of 91 compound categories encoding shape, color, and orientation. This ontology is derived from an empirical analysis of six major VRC providers, covering both the visual primitives present in images and the semantic attributes referenced in their queries, and thus reflects the common structure of mainstream VRC deployments rather than arbitrary design choices.

The detector demonstrates strong cross-platform generalization (Table~\ref{tab:yolo_metrics}), indicating that the ontology captures provider-invariant features rather than overfitting to specific templates. When future VRCs introduce new visual relations (e.g., unseen shapes, colors, relations), the impact is confined to the perception layer: objects outside the ontology are conservatively filtered out or misclassified, yielding smaller candidate sets rather than hallucinated targets. Attack success rates degrade due to perception errors, while the downstream perception–reasoning pipeline and reasoning logic remain unchanged.

\begin{table}[h]
\centering
\scriptsize
\caption{\textbf{Detector validation summary.}}
\label{tab:yolo_metrics}
\begin{threeparttable}
\resizebox{1\linewidth}{!}{
\begin{tabular}{cccccc}
\toprule
\multicolumn{6}{c}{\textbf{(A) Overall COCO-style metrics}} \\
\cmidrule{1-6}
 Precision & Recall & mAP@50 & mAP@50:95 & F1\textsubscript{opt} & Best epoch \\
 \cellcolor{green!20}0.998 & 0.956 & 0.984 & 0.919 & 0.968 & \cellcolor{green!20}178 \\
\midrule
\multicolumn{6}{c}{\textbf{(B) Stratified AP (by size/attribute)}} \\
\cmidrule{1-6}
AP\textsubscript{s} & AP\textsubscript{m} & AP\textsubscript{l} & AR@100 & TP/img & FP/img \\
 0.912 & 0.944 & \cellcolor{green!20}0.963 & 0.931 & 3.52 & 0.11 \\
\midrule
\multicolumn{6}{c}{\textbf{(C) Per-platform mAP}} \\
\cmidrule{1-6}
VTT & Geetest & NetEase & DingX & Shumei & Xiaodun \\
\rowcolor{blue!5}
\multicolumn{6}{c}{\emph{mAP@0.5}} \\
\cellcolor{orange!20}0.999 & 0.998 & 0.998 & \cellcolor{green!20}1.000 & \cellcolor{red!20}0.950 & 0.958 \\
\rowcolor{green!5}
\multicolumn{6}{c}{\emph{mAP@0.5:0.95}} \\
\cellcolor{green!20}0.965 & \cellcolor{red!20}0.907 & 0.910 & 0.951 & 0.922 & \cellcolor{orange!20}0.957 \\
\midrule
\multicolumn{6}{c}{\textbf{(D) Runtime Statistics}} \\
\cmidrule{1-6}
 \# Params & FLOPs & Lat. & Thrpt. & NMS & Mean$\pm$Std \\
 25.6M & 72.4G & 7.8ms & 128/s & 0.9ms & \cellcolor{green!20}$0.932\pm0.009$ \\
\bottomrule
\end{tabular}
}
\begin{tablenotes}
       \footnotesize
       \item[] \emph{AP\textsubscript{s}, AP\textsubscript{m}, AP\textsubscript{l}} denote average precision for small, medium, and large objects; 
\emph{FLOPs} indicates floating point operations at 640$\times$640 resolution; 
\emph{Lat.} and \emph{Thrpt.} refer to latency and throughput; 
\emph{NMS} stands for non-maximum suppression.
\end{tablenotes}
\end{threeparttable}
\end{table}

\begin{table*}[t]
\centering
\renewcommand{\arraystretch}{1.1}
\scriptsize
\caption{\textbf{Mean response time} (seconds) across CAPTCHA platforms. Reported as mean $\pm$ standard deviation.}
\label{tab:response_time_summary}
\vspace{6pt}
\begin{tabularx}{\linewidth}{>{\columncolor{kimicolor}}l|YYYYYY}
\toprule
\multicolumn{1}c{\textbf{Solver}} & \textbf{VTT} & \textbf{Geetest} & \textbf{NetEase} & \textbf{DingXiang} & \textbf{Shumei} & \textbf{Xiaodun} \\
\midrule
Holistic & $0.05 \ (\pm 0.01)$ & $0.05 \ (\pm 0.01)$ & $0.05 \ (\pm 0.01)$ & $0.05 \ (\pm 0.01)$ & $0.05 \ (\pm 0.01)$ & $0.05 \ (\pm 0.01)$ \\
GraphNet & $0.20 \ (\pm 0.05)$ & $0.20 \ (\pm 0.05)$ & $0.20 \ (\pm 0.05)$ & $0.20 \ (\pm 0.05)$ & $0.20 \ (\pm 0.05)$ & $0.20 \ (\pm 0.05)$ \\
Oedipus  & $3.10 \ (\pm 0.40)$ & $3.20 \ (\pm 0.45)$ & $3.30 \ (\pm 0.42)$ & $2.90 \ (\pm 0.38)$ & $3.50 \ (\pm 0.44)$ & $3.00 \ (\pm 0.41)$ \\
\cmidrule{1-7}
VIPER (gpt-4o)   & $5.31 \ (\pm 1.39)$ & $5.10 \ (\pm 1.44)$ & $6.17 \ (\pm 1.53)$ & $6.06 \ (\pm 0.95)$ & $5.54 \ (\pm 1.12)$ & $4.42 \ (\pm 0.89)$ \\
VIPER (deepseek) & $10.20 \ (\pm 1.12)$ & $9.80 \ (\pm 1.35)$ & $9.90 \ (\pm 1.25)$ & $9.10 \ (\pm 1.40)$ & $8.50 \ (\pm 1.10)$ & $8.70 \ (\pm 1.30)$ \\
VIPER (kimi)     & $9.80 \ (\pm 1.43)$  & $11.40 \ (\pm 1.77)$ & $9.58 \ (\pm 3.02)$  & $7.20 \ (\pm 0.20)$ & $11.37 \ (\pm 2.45)$ & $9.27 \ (\pm 2.00)$ \\
VIPER (grok-2)   & $8.57 \ (\pm 1.11)$  & $8.97 \ (\pm 1.10)$  & $8.33 \ (\pm 0.35)$  & $7.00 \ (\pm 2.07)$ & $6.67 \ (\pm 1.20)$ & $6.50 \ (\pm 1.85)$ \\
\cmidrule{1-6}
\textbf{VIPER (overall)} & 
\textbf{8.97} & 
\textbf{8.82} & 
\textbf{8.99} & 
\textbf{7.84} & 
\textbf{8.52} & 
\textbf{7.97} \\
\cmidrule{1-7}
Human baseline   & $7.40 \ (\pm 2.10)$ & $7.80 \ (\pm 2.30)$ & $6.90 \ (\pm 1.90)$ & $6.60 \ (\pm 1.80)$ & $6.10 \ (\pm 1.70)$ & $7.20 \ (\pm 2.00)$ \\
\bottomrule
\end{tabularx}
\end{table*}

\section{Defenses Against VRC Attacks}
\label{sec-defense}

\noindent\textbf{Defense approach.}
The key weakness exploited by VRC solvers is their reliance on narrow linguistic templates. Once a model has seen enough instructions phrased in a fixed style, it can map these prompts to a predictable visual grounding pattern, thereby shortcutting the intended reasoning process. To counter this, we propose \textit{Template-Space Randomization (TSR)}, a lightweight defense that enlarges the linguistic surface of the challenge without altering its semantics. TSR diversifies prompts along three axes: (i) \emph{synonym substitution} for object attributes (e.g., replacing ``red'' with ``scarlet''), (ii) \emph{relation rewording and polarity flips} (e.g., ``to the left of'' $\leftrightarrow$ ``not to the right of''), and (iii) \emph{optional indirection}, where the user must resolve an extra reference (e.g., ``click the item that shares color with the cone to the left of \texttt{T}'').


\smallskip
\noindent\textbf{Evaluation.}  
We conducted a preliminary evaluation of TSR using a subset of 100 original VRC images from our benchmark. For each image, we instantiated six randomized variants of the question prompts, corresponding to three transformation dimensions, synonym substitution for object attributes (S), relation rewording and polarity flips (R), and optional indirection (I), along with their pairwise (S+R, S+I) and full (S+R+I) combinations. This yielded a controlled benchmark of 600 test queries. Importantly, the visual content and ground-truth answers remained unchanged; only the linguistic surface form of the questions was varied through TSR. We then tested four representative solvers, \textsc{ViPer} (GPT-4o), Oedipus, GraphNet, and Holistic, on these challenges to obtain an initial assessment of robustness under TSR. While limited in scale, these results provide early evidence that linguistic surface variability alone can measurably degrade solver performance without altering task semantics.

\begin{table}[!htbp]
\centering
\scriptsize
\caption{\textbf{TSR effectiveness (success rate, \%).} Averages across six platforms. S = synonyms; R = relation rewording/polarity; I = indirection.}
\label{tab:tsr_success}
\vspace{4pt}
\begin{tabularx}{\columnwidth}{>{\columncolor{kimicolor}}l|c|cccccc}
\toprule
\multicolumn{1}{c}{\textbf{Solvers}} & \multicolumn{1}{c}{\textbf{Original}} & \textbf{S} & \textbf{R} & \textbf{I} & \textbf{S+R} & \textbf{S+I} & \textbf{S+R+I} \\
\midrule
\textsc{ViPer} (GPT-4o) & 93.60 & \cellcolor{green!20}91.20 & 90.10 & 89.50 & 86.40 & \cellcolor{orange!20}86.20 & \cellcolor{red!20}85.20 \\
Oedipus                 & 65.80 & \cellcolor{green!20}61.00 & 56.90 & 58.70 & \cellcolor{orange!20}49.10 & 50.80 & \cellcolor{red!20}45.30 \\
GraphNet                & 83.20 & \cellcolor{green!20}81.00 & 79.80 & 78.40 & \cellcolor{orange!20}71.60 & 74.10 & \cellcolor{red!20}68.40 \\
Holistic                & 89.40 & 87.70 & \cellcolor{green!20}88.30 & 87.90 & 86.90 & \cellcolor{orange!20}86.50 & \cellcolor{red!20}84.90 \\
\bottomrule
\end{tabularx}
\vspace{-0.3em}
\end{table}

\smallskip
\noindent\textbf{Findings.}  
TSR reduces solver performance to varying degrees (Table~\ref{tab:tsr_success}). \textsc{ViPer} (GPT-4o) remains relatively resilient, with only a modest decline from 93.60\% to 85.20\% under the full S+R+I setting. Holistic displays a similar level of stability, decreasing slightly from 89.40\% to 84.90\%. In contrast, Oedipus is more vulnerable, dropping from 65.80\% to 45.30\%, while GraphNet falls from 83.20\% to 68.40\%. The results indicate that even when the underlying images and correct answers remain unchanged, simple linguistic variability in the question prompts can significantly affect solver performance. In other words, robustness to surface-level linguistic variation remains an open challenge, highlighting that solvers often rely on brittle language–vision alignments rather than deeper semantic understanding.

\section{Limitations and Future Work}
First, our evaluation relies on advanced LLMs to establish an upper bound on the capability of the proposed framework. We did not systematically study smaller or lightweight LLM backbones, and thus cannot fully disentangle the contribution of \textsc{ViPer}'s modular design from the raw reasoning capacity of advanced models. Understanding how performance degrades with reduced model scale for resource-constrained settings, remains an important direction for future work.

Second, another limitation of this study is the lack of a systematic computational cost analysis. \textsc{ViPer} employs a multi-stage pipeline that combines a multi-object detector with LLM-based reasoning, which inevitably incurs non-trivial inference latency and API cost when using LLMs. Our evaluation focuses on accuracy and robustness, and does not quantify cost--performance trade-offs. A cost-sensitive analysis is therefore left to future work.

Third, \textsc{ViPer} focuses on a specific category of challenge-based CAPTCHAs, namely VRCs. While VRCs represent one of the most challenging and widely deployed CAPTCHA classes in practice, our findings do not directly generalize to other paradigms such as behavioral or interaction-based systems. Extending the perception--reasoning framework to broader CAPTCHA families, or understanding the limits of such extensions, remains an open research problem.

Finally, our experiments assess the effectiveness of the proposed TSR defense only against automated solvers, and we did not conduct human-subject studies to verify whether the obfuscated or redirected language remains intuitive and solvable for legitimate users. As a result, the practical usability of TSR cannot yet be fully established, and evaluating human success rates and cognitive load under such defenses remains a direction for future work.

\section{Related Work}
\label{apd-rw}

\noindent\textbf{CAPTCHA evolution.}
The design of CAPTCHAs has long reflected an arms race between defensive schemes and automated solvers. Early text-based CAPTCHAs exploited weaknesses in OCR by introducing distortions, clutter, and background noise~\cite{gao2016simple,bursztein2014end}. As vision models improved, image-based CAPTCHAs shifted the challenge to object recognition and semantic understanding, but growing model capabilities eroded their robustness while exacerbating usability issues~\cite{zhang2025privcaptcha,guerar2021gotta,alqahtani2020image}. To balance security and user experience, interactive CAPTCHAs were introduced, relying on higher-level reasoning tasks thought to be beyond automation~\cite{li2021vrcaptcha,teoh2024phishdecloaker,searles2023empirical,gangwal2025swiss}. More recently, systems such as reCAPTCHA v3~\cite{google2018recaptcha} abandoned puzzles, replacing them with risk-based, invisible checks that analyze behavioral biometrics (e.g., cursor dynamics, typing cadence) to assign trust scores transparently. Parallel research has also explored adversarial CAPTCHAs~\cite{shi2021adversarial,wang2023improving,zhang2022counteracting}, which embed human-imperceptible perturbations to mislead automated solvers.

\smallskip
\noindent\textbf{Attacks on CAPTCHAs.}
A large body of work demonstrated the vulnerability of text CAPTCHA schemes to automated attacks. Gao et al.~\cite{gao2017security} showed that Microsoft’s two-layer CAPTCHA could be broken with 44.6\% accuracy in 9.05 seconds, while Tang et al.~\cite{tang2018captcha} developed a fast deep learning-based attack on text CAPTCHAs. Subsequent work employed CAPTCHA synthesizers to generate synthetic challenges for training baseline solvers~\cite{ye2018yet}. More advanced approaches exploited cycle-consistent GANs~\cite{li2021end}, masked autoencoders~\cite{zhao2023geesolver}, CNN architectures~\cite{zi2019end, sengul2025recognition}, and unsupervised learning~\cite{tian2020generic}, establishing end-to-end solvers for diverse CAPTCHA formats. Even segmentation-resistant designs have been shown to remain vulnerable~\cite{wang2023experimental}.

Beyond text-based schemes, image-based CAPTCHAs are also susceptible. Researchers have bypassed slider CAPTCHAs by combining visual alignment techniques (e.g., template matching, edge detection) with imitation of human-like mouse dynamics~\cite{dai2024c,chang2024robustness}. Rotation-based CAPTCHAs have similarly been compromised using deep learning models trained on augmented datasets or classification across discrete angle bins~\cite{gossweiler2009s,tang2018research}.

\smallskip
\noindent\textbf{LLM-based visual reasoning.}  
Recent progress extends visual reasoning by coupling structured perception with the compositional abilities of LLMs~\cite{brown2020language,chowdhery2023palm,touvron2023llama}. Early tasks such as visual question answering (VQA)~\cite{antol2015vqa,goyal2017making} and visual entailment~\cite{xie2019snlive} highlight the need to align linguistic queries with visual semantics. Knowledge-augmented benchmarks (e.g., OK-VQA, A-OKVQA) further reveal the importance of external grounding~\cite{marino2019okv,schwenk2022okvqa}. Recent vision–language models like Flamingo~\cite{alayrac2022flamingo}, PaLM-E~\cite{driess2023palm}, and BLIP-2~\cite{li2023blip} show that LLM reasoning can be paired with pretrained perception modules, but systematic methods for robust integration remain open~\cite{yu2025survey,chang2024survey}.

\section{Conclusion}

We presented \textsc{ViPer}, a perception–reasoning attack framework that integrates multi-object detection with adaptive LLM inference. By decoupling visual grounding from symbolic reasoning, \textsc{ViPer} generalizes across heterogeneous VRCs without retraining. We evaluated it under four protocols: (i) component ablation, showing monotonic gains up to 93.2\% accuracy; (ii) benchmark comparison, outperforming GraphNet (83.2\%), OEDIPUS (65.8\%), and the Holistic baseline (89.5\%); (iii) LLM backbone swap, maintaining robust performance ($>$90\%) across GPT-4o, DeepSeek, Kimi, and Grok; and (iv) response-time analysis, achieving latency comparable to human solving. Overall, \textsc{ViPer} matches human accuracy on most datasets and establishes a new upper bound for VRC solvers, highlighting the urgent need for human-solvable yet machine-resilient VRCs defenses.

\section*{Ethical Considerations}
This work studies the security of VRCs, a class of widely deployed human-interaction proofs. We recognize that research on automated CAPTCHA solving is \emph{dual-use}: it informs the design of more robust, human-solvable defences but may also be misused to bypass safeguards. Our goal is to support responsible mitigation rather than facilitate abuse; we do not provide deployment-ready attack tools, nor do we automate large-scale abuse.

\smallskip
\noindent\textbf{Stakeholders.}  
We identify several stakeholder groups potentially impacted.  
(1) \emph{Service providers} (e.g., VRC vendors and websites relying on them) may face increased security risks if vulnerabilities are exploited.  
(2) \emph{End users} may be indirectly affected if weakened CAPTCHAs lead to higher abuse rates (e.g., spam or fraud).  
(3) \emph{Researchers and system designers} benefit from clearer understanding of the limits of current VRC defenses.  
(4) \emph{The broader Internet ecosystem}, including communities often overlooked in security research, may be impacted by shifts in defensive strategies.  
(5) \emph{The authors themselves}, who must balance scientific disclosure against the risk of enabling misuse.

\smallskip
\noindent\textbf{Potential impacts and dual use.}  
The primary risk of this work is that a generalized VRC solver could be abused to circumvent security mechanisms, leading to spam, fraud, or denial-of-service amplification. At the same time, understanding these vulnerabilities has positive impacts: it enables vendors to assess whether VRCs still provide meaningful security guarantees and informs the development of human-solvable yet machine-resistant challenges. We believe the benefits of transparent analysis outweigh the risks, provided appropriate safeguards are applied.

\smallskip
\noindent\textbf{Mitigations and safeguards.}  
First, all experiments were conducted \emph{offline}, without interacting with production systems or live user traffic. Benchmarks were collected using public demo portals or self-registered API keys, avoiding disruption to real services. Second, we practised \emph{responsible disclosure}: findings were communicated to all VRC providers prior to publication. Third, we support reproducibility but the released artifacts are limited to research code and sanitized datasets; sensitive identifiers are removed. Access is restricted to research purposes.

\smallskip
\noindent\textbf{Human subjects.}  
Our human baseline study involved minimal risk and no deception. Participants were informed of the task purpose, provided consent, and could withdraw at any time. No personally identifiable information was collected, and all results were aggregated. 

\smallskip
\noindent\textbf{Data collection.}
Our study uses only public VRC challenges collected under normal usage constraints, without bypassing access controls or interacting with backend systems. We do not collect personal data or user identifiers. All datasets are curated using consistent protocols to ensure experimental integrity and reproducibility.


\section*{Open Science}
To enable transparency and reproducibility, we release our source code and benchmark resources at  
\url{https://zenodo.org/records/17971722}. The package includes the full \textsc{ViPer} pipeline (model configurations, preprocessing scripts, and ablation modules), as well as cleaned, standardized versions of the six VRC datasets (VTT, Geetest, NetEase, DingXiang, Shumei, Xiaodun). We also provide unified evaluation protocols based on COCO-style metrics. The release aligns with the broader open-science policies.

\bibliographystyle{IEEEtran}
\bibliography{ref(short)}

\begin{thebibliography}{10}
\providecommand{\url}[1]{#1}
\csname url@samestyle\endcsname
\providecommand{\newblock}{\relax}
\providecommand{\bibinfo}[2]{#2}
\providecommand{\BIBentrySTDinterwordspacing}{\spaceskip=0pt\relax}
\providecommand{\BIBentryALTinterwordstretchfactor}{4}
\providecommand{\BIBentryALTinterwordspacing}{\spaceskip=\fontdimen2\font plus
\BIBentryALTinterwordstretchfactor\fontdimen3\font minus \fontdimen4\font\relax}
\providecommand{\BIBforeignlanguage}[2]{{%
\expandafter\ifx\csname l@#1\endcsname\relax
\typeout{** WARNING: IEEEtran.bst: No hyphenation pattern has been}%
\typeout{** loaded for the language `#1'. Using the pattern for}%
\typeout{** the default language instead.}%
\else
\language=\csname l@#1\endcsname
\fi
#2}}
\providecommand{\BIBdecl}{\relax}
\BIBdecl

\bibitem{uzun2018rtcaptcha}
E.~Uzun \emph{et~al.}, ``rtcaptcha: A real-time {CAPTCHA} based liveness detection system.'' \emph{NDSS}, 2018.

\bibitem{tang2018captcha}
M.~Tang, H.~Gao \emph{et~al.}, ``Research on deep learning techniques in breaking text-based {CAPTCHA}s and designing image-based {CAPTCHA},'' \emph{TIFS}, 2018.

\bibitem{shi2020text}
C.~Shi, S.~Ji \emph{et~al.}, ``Text captcha is dead? a large scale deployment and empirical study,'' \emph{CCS}, 2020.

\bibitem{guerar2021gotta}
M.~Guerar \emph{et~al.}, ``Gotta {CAPTCHA}’em all: A survey of 20 years of the human-or-computer dilemma,'' \emph{ACM Computing Surveys}, 2021.

\bibitem{wang2023experimental}
P.~Wang, H.~Gao \emph{et~al.}, ``An experimental investigation of text-based {CAPTCHA} attacks and their robustness,'' \emph{ACM Computing Surveys}, 2023.

\bibitem{ousat2024matter}
B.~Ousat, E.~Schafir \emph{et~al.}, ``The matter of captchas: An analysis of a brittle security feature on the modern web,'' in \emph{WWW}, 2024.

\bibitem{wang2018captcha}
H.~Wang, F.~Zheng \emph{et~al.}, ``A captcha design based on visual reasoning,'' \emph{ICASSP}, 2018.

\bibitem{gao2021research}
Y.~Gao \emph{et~al.}, ``Research on the security of visual reasoning {CAPTCHA},'' \emph{USENIX Sec}, 2021.

\bibitem{wang2023extended}
P.~Wang \emph{et~al.}, ``Extended research on the security of visual reasoning {CAPTCHA},'' \emph{TDSC}, 2023.

\bibitem{ding2025illusioncaptcha}
Z.~Ding \emph{et~al.}, ``{IllusionCAPTCHA}: A {CAPTCHA} based on visual illusion,'' in \emph{WWW}, 2025.

\bibitem{vtt}
{Yiyepianzhou et al.}, ``{VTT CAPTCHA Dataset (ViRC)},'' \url{https://cloud.tencent.com/product/captcha}, 2024.

\bibitem{geetest}
{GeeTest}, ``{GeeTest CAPTCHA Platform},'' \url{https://www.geetest.com/en}, 2025.

\bibitem{builtwith_recaptcha}
{BuiltWith}, ``Captcha usage statistics on the entire internet,'' \url{https://trends.builtwith.com/widgets/captcha/traffic/Entire-Internet}, 2024.

\bibitem{google2018recaptcha}
\BIBentryALTinterwordspacing
{Google}, ``recaptcha v3 documentation,'' Webpage, 2018. [Online]. Available: \url{https://developers.google.com/recaptcha/docs/v3}
\BIBentrySTDinterwordspacing

\bibitem{george2017generative}
D.~George, W.~Lehrach \emph{et~al.}, ``A generative vision model that trains with high data efficiency and breaks text-based captchas,'' \emph{Science}, 2017.

\bibitem{plesner2024breaking}
A.~Plesner, T.~Vontobel, and R.~Wattenhofer, ``Breaking recaptchav2,'' in \emph{COMPSAC}, 2024.

\bibitem{teohcaptchas}
X.~Teoh \emph{et~al.}, ``Are captchas still bot-hard? generalized visual captcha solving with agentic vision language model,'' \emph{USENIX Sec}, 2025.

\bibitem{xu2023vrc}
B.~Xu and H.~Wang, ``Vrc-graphnet: A graph neural network-based reasoning framework for attacking visual reasoning {CAPTCHA}s,'' \emph{SecureComm}, 2023.

\bibitem{achiam2023gpt}
J.~Achiam \emph{et~al.}, ``{GPT}-4 technical report,'' \emph{arXiv preprint arXiv:2303.08774}, 2023.

\bibitem{grok}
{xAI}, ``{Grok by xAI},'' \url{https://x.ai}, 2024.

\bibitem{xu2023multimodal}
P.~Xu, X.~Zhu, and D.~A. Clifton, ``Multimodal learning with transformers: A survey,'' \emph{TPAMI}, 2023.

\bibitem{zhou2024large}
H.~Zhou \emph{et~al.}, ``Large language model ({LLM}) for telecommunications: A comprehensive survey on principles, key techniques, and opportunities,'' \emph{IEEE Communications Surveys \& Tutorials}, 2024.

\bibitem{yu2025survey}
M.~Yu \emph{et~al.}, ``A survey on trustworthy {LLM} agents: Threats and countermeasures,'' \emph{KDD}, 2025.

\bibitem{chang2024survey}
Y.~Chang, X.~Wang \emph{et~al.}, ``A survey on evaluation of large language models,'' \emph{TIST}, 2024.

\bibitem{yang2024harnessing}
J.~Yang \emph{et~al.}, ``Harnessing the power of {LLM}s in practice: A survey on chatgpt,'' \emph{TKDD}, 2024.

\bibitem{deng2024Oedipus}
J.~Deng \emph{et~al.}, ``{OEDIPUS}: {LLM}-enhanced reasoning {CAPTCHA} solver,'' \emph{USENIX Sec}, 2024.

\bibitem{xiaodun}
{Xiaodun AI}, ``{Xiaodun {CAPTCHA} Platform},'' \url{https://xiaodun.com/}, 2025.

\bibitem{netease}
{NetEase}, ``{NetEase CAPTCHA Service},'' \url{https://dun.163.com/trial/space-inference}, 2025.

\bibitem{dingxiang}
{Dingxiang Technologies}, ``{Dingxiang CAPTCHA Platform},'' \url{https://www.dun.163.com}, 2025.

\bibitem{shumei}
{Shumei Tech}, ``{Shumei CAPTCHA Service},'' \url{https://www.ishumei.com/trial/captcha.html}, 2025.

\bibitem{yolov11}
U.~Team, ``Yolov11 (community prototype),'' \url{https://github.com/ultralytics/ultralytics}, 2024.

\bibitem{gao2016simple}
H.~Gao, J.~Yan \emph{et~al.}, ``A simple generic attack on text captchas.'' in \emph{NDSS}, 2016.

\bibitem{bursztein2014end}
E.~Bursztein, J.~Aigrain, A.~Moscicki, and J.~C. Mitchell, ``The end is nigh: Generic solving of text-based $\{$CAPTCHAs$\}$,'' in \emph{USENIX WOOT}, 2014.

\bibitem{zhang2025privcaptcha}
S.~Zhang, X.~Yi \emph{et~al.}, ``Privcaptcha: Interactive captcha to facilitate effective comprehension of app privacy policy,'' in \emph{CHI}, 2025.

\bibitem{alqahtani2020image}
F.~Alqahtani \emph{et~al.}, ``Is image-based {CAPTCHA} secure against attacks based on machine learning? an experimental study,'' \emph{Computers\&Security}, 2020.

\bibitem{li2021vrcaptcha}
X.~Li, Y.~Chen \emph{et~al.}, ``vrcaptcha: exploring captcha designs in virtual reality,'' in \emph{CHI}, 2021.

\bibitem{teoh2024phishdecloaker}
X.~Teoh \emph{et~al.}, ``Detecting {CAPTCHA-cloaked} phishing websites via hybrid vision-based interactive models,'' \emph{USENIX Sec}, 2024.

\bibitem{searles2023empirical}
A.~Searles \emph{et~al.}, ``An empirical study \& evaluation of modern $\{$CAPTCHAs$\}$,'' \emph{USENIX Sec}, 2023.

\bibitem{gangwal2025swiss}
A.~Gangwal, P.~S. Reddy, and C.~y. Sagar, ``Swiss cheese captcha: A novel multi-barrier mechanism for bot detection,'' \emph{SAC}, 2025.

\bibitem{shi2021adversarial}
C.~Shi, X.~Xu, S.~Ji \emph{et~al.}, ``Adversarial captchas,'' \emph{IEEE Transactions on Cybernetics}, 2021.

\bibitem{wang2023improving}
P.~Wang \emph{et~al.}, ``Improving the security of audio captchas with adversarial examples,'' \emph{TDSC}, 2023.

\bibitem{zhang2022counteracting}
N.~Zhang \emph{et~al.}, ``Counteracting dark web text-based {CAPTCHA} with generative adversarial learning for proactive cyber threat intelligence,'' \emph{TMIS}, 2022.

\bibitem{gao2017security}
H.~Gao, M.~Tang \emph{et~al.}, ``Research on the security of microsoft's two-layer {CAPTCHA},'' \emph{TIFS}, 2017.

\bibitem{ye2018yet}
G.~Ye, Z.~Tang \emph{et~al.}, ``Yet another text captcha solver: A generative adversarial network based approach,'' in \emph{CCS}, 2018.

\bibitem{li2021end}
C.~Li, X.~Chen \emph{et~al.}, ``End-to-end attack on text-based {CAPTCHA}s based on cycle-consistent generative adversarial network,'' \emph{Neurocomputing}, 2021.

\bibitem{zhao2023geesolver}
R.~Zhao \emph{et~al.}, ``Geesolver: A generic, efficient, and effortless solver with self-supervised learning for breaking text {CAPTCHA}s,'' \emph{S\&P}, 2023.

\bibitem{zi2019end}
Y.~Zi, H.~Gao, Z.~Cheng, and Y.~Liu, ``An end-to-end attack on text {CAPTCHA}s,'' \emph{TIFS}, 2019.

\bibitem{sengul2025recognition}
B.~B. Seng{\"u}l and M.~B. Alver, ``Recognition of multiple captcha types: Using cnn and combined dataset,'' \emph{SİU}, 2025.

\bibitem{tian2020generic}
S.~Tian \emph{et~al.}, ``A generic solver combining unsupervised learning and representation learning for breaking text-based {CAPTCHA}s,'' in \emph{WWW}, 2020.

\bibitem{dai2024c}
H.~Dai~Nguyen \emph{et~al.}, ``C-frame: Characterizing and measuring in-the-wild captcha attacks,'' \emph{S\&P}, 2024.

\bibitem{chang2024robustness}
G.~Chang \emph{et~al.}, ``The robustness of behavior-verification-based slider {CAPTCHA}s,'' 2024.

\bibitem{gossweiler2009s}
R.~Gossweiler, M.~Kamvar, and S.~Baluja, ``What's up {CAPTCHA}? a {CAPTCHA} based on image orientation,'' in \emph{WWW}, 2009.

\bibitem{tang2018research}
M.~Tang, H.~Gao \emph{et~al.}, ``Research on deep learning techniques in breaking text-based {CAPTCHA}s and designing image-based {CAPTCHA},'' \emph{TIFS}, 2018.

\bibitem{brown2020language}
T.~Brown, B.~Mann \emph{et~al.}, ``Language models are few-shot learners,'' \emph{NeurIPS}, 2020.

\bibitem{chowdhery2023palm}
A.~Chowdhery, S.~Narang \emph{et~al.}, ``Palm: Scaling language modeling with pathways,'' \emph{JMLR}, 2023.

\bibitem{touvron2023llama}
H.~Touvron, T.~Lavril \emph{et~al.}, ``Llama: Open and efficient foundation language models,'' \emph{arXiv preprint arXiv:2302.13971}, 2023.

\bibitem{antol2015vqa}
S.~Antol, A.~Agrawal \emph{et~al.}, ``Vqa: Visual question answering,'' \emph{ICCV}, 2015.

\bibitem{goyal2017making}
Y.~Goyal, T.~Khot \emph{et~al.}, ``Making the v in vqa matter: Elevating the role of image understanding in visual question answering,'' \emph{CVPR}, 2017.

\bibitem{xie2019snlive}
N.~Xie \emph{et~al.}, ``Visual entailment: A novel task for fine-grained image understanding (snli-ve),'' \emph{arXiv preprint arXiv:1901.06706}, 2019.

\bibitem{marino2019okv}
K.~Marino, M.~Rastegari, A.~Farhadi, and R.~Mottaghi, ``Ok-vqa: A visual question answering benchmark requiring external knowledge,'' \emph{CVPR}, 2019.

\bibitem{schwenk2022okvqa}
D.~Schwenk, A.~Khandelwal, C.~Clark \emph{et~al.}, ``A-okvqa: A benchmark for visual question answering using world knowledge,'' \emph{ECCV}, 2022.

\bibitem{alayrac2022flamingo}
J.-B. Alayrac \emph{et~al.}, ``Flamingo: a visual language model for few-shot learning,'' \emph{NeurIPS}, 2022.

\bibitem{driess2023palm}
D.~Driess, F.~Xia \emph{et~al.}, ``Palm-e: an embodied multimodal language model,'' in \emph{ICML}, 2023.

\bibitem{li2023blip}
J.~Li, D.~Li \emph{et~al.}, ``Blip-2: Bootstrapping language-image pre-training with frozen image encoders and large language models,'' in \emph{ICML}, 2023.

\bibitem{von2004telling}
L.~von Ahn, M.~Blum \emph{et~al.}, ``Telling humans and computers apart automatically,'' \emph{CACM}, 2004.

\bibitem{mori2003recognizing}
G.~Mori and J.~Malik, ``Recognizing objects in adversarial clutter: Breaking a visual {CAPTCHA},'' \emph{CVPR}, 2003.

\bibitem{gossweiler2009whatsup}
R.~Gossweiler, M.~Kamvar, and S.~Baluja, ``What's up {CAPTCHA}? a {CAPTCHA} based on image orientation,'' in \emph{WWW}, 2009.

\bibitem{chellapilla2005designing}
K.~Chellapilla, K.~Larson, P.~Simard, and M.~Czerwinski, ``Designing human friendly human interaction proofs ({HIP}s),'' in \emph{CHI}, 2005.

\bibitem{von2008recaptcha}
L.~von Ahn, B.~Maurer, C.~McMillen \emph{et~al.}, ``{reCAPTCHA}: Human-based character recognition via web security measures,'' \emph{Science}, 2008.

\bibitem{googleRecaptcha}
{Google}, ``{reCAPTCHA v2},'' \url{https://cloud.google.com/security/products/recaptcha}, 2014.

\bibitem{hossen2020object}
M.~Hossen \emph{et~al.}, ``An object detection based solver for {Google’s} image {reCAPTCHA} v2,'' \emph{RAID}, 2020.

\bibitem{li2025mi}
J.~Li, L.~Fu \emph{et~al.}, ``Mi-captcha: Enhance the security of captcha using mooney images,'' \emph{AAAI}, 2025.

\end{thebibliography}

\appendix

\section{Technique Supplementary }
\label{apd-tech}

\subsection{CAPTCHAs}

\begin{figure}[t]
    \centering
    \includegraphics[width=0.85\columnwidth]
    {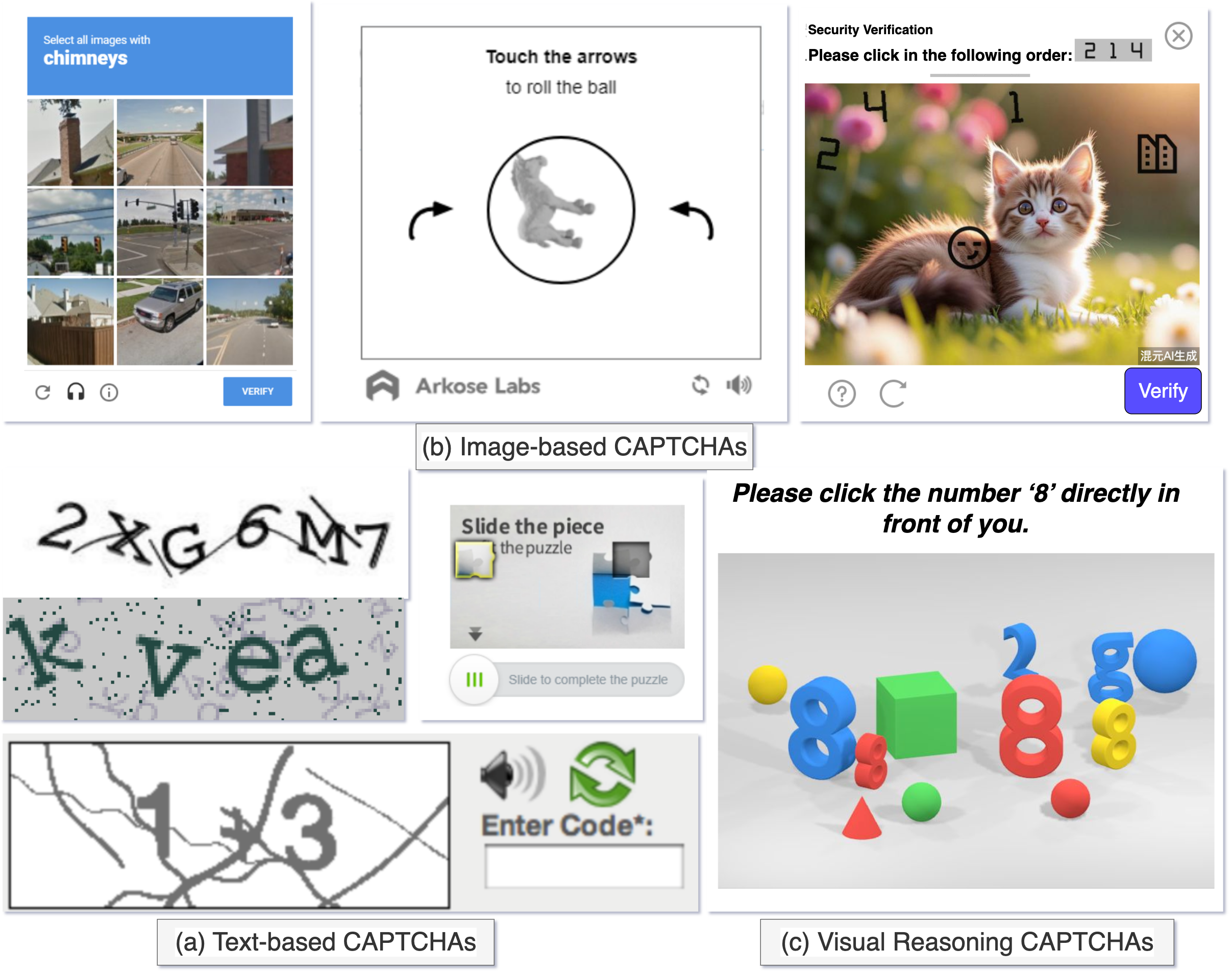}
    \vspace{-0.2in}
    \caption{Examples of \textbf{CAPTCHA types}.}
    \label{fig:categories}
    \vspace{-0.2in}
\end{figure}

\noindent\textbf{Text-based CAPTCHAs} (Fig.\ref{fig:categories}(a)). These are the earliest and most widely adopted HIPs. Their design principle is to distort or obfuscate characters such that they remain legible to humans but difficult for optical character recognition (OCR). Schemes such as Input CAPTCHA~\cite{von2004telling}, Gimpy~\cite{mori2003recognizing}, and PessimalPrint~\cite{gossweiler2009whatsup} increase robustness by introducing background clutter, overlapping characters, or degraded font rendering~\cite{chellapilla2005designing, von2008recaptcha, ousat2024matter}.

\smallskip
\noindent\textbf{Image-based CAPTCHAs} (Fig.\ref{fig:categories}(b)). These exploit human visual perception to differentiate humans from bots. They offer richer interaction modes and higher resilience to OCR attacks than text-based schemes~\cite{googleRecaptcha,hossen2020object,li2025mi}. We categorise designs into four interaction modes:

\begin{packeditemize}

\item \textit{Single click:} Users need to click on a specific area of a single image (e.g., ``\textit{click on the mountain}'') or identify particular objects across multiple sub-images (``\textit{select all images containing cats}''). This type is widely used (e.g., Google reCAPTCHA v2). However, deep learning can be trained to solve click-based CAPTCHAs, and identify common object categories like traffic lights, buses, or crosswalks.

\item \textit{Sequential clicks:}
Users must click on specified symbols, characters, or objects in a particular sequence (e.g., ``\textit{clicking Chinese characters in order}''). This increases the complexity for bots, as it combines object recognition with memory and sequence tracking.

\item \textit{Slider:} Slider CAPTCHAs require users to drag a puzzle piece or slider to fit into a target position within an image. They include behavioral analysis (e.g., ``\textit{tracking mouse movement patterns}'') to further distinguish human input from scripts.

\item \textit{Rotation-based CAPTCHAs:}
These CAPTCHAs ask users to rotate images to the correct orientation (e.g., ``\textit{making an upside-down building upright}''). Such tasks are simple for humans but non-trivial for bots, especially without access to robust orientation classifiers.

\end{packeditemize}

\noindent\textbf{Visual reasoning CAPTCHAs} (Fig.\ref{fig:categories}(c)).
VRCs require users to perform multi-step visual reasoning, i.e., spatial, comparative, and logical inference over complex scenes~\cite{ding2025illusioncaptcha}. Each challenge presents a scene containing multiple objects (e.g., geometric shapes, letters, digits) along with a natural language instruction that encodes the query logic. Solving VRCs involves interpreting the instruction, understanding relationships between objects (such as ``\textit{to the left of}'', ``\textit{same color as}'', or ``\textit{the largest}''), and accurately locating the correct target region.

\begin{table}[!]
\renewcommand{\arraystretch}{1} 
\setlength{\tabcolsep}{3pt}     
\centering
\scriptsize
\caption{\textbf{Zero-shot performance of GPT-4o} on seven widely used CAPTCHA types.}
\label{tab:captcha_types_summary}
\vspace{3pt}
\begin{tabularx}{\linewidth}{>{\columncolor{kimicolor}}c|c|cc}
\toprule
\multicolumn{1}{c}{\textbf{Type}} & \multicolumn{1}{c}{\textbf{Test cases}} & \textbf{Accuracy (\%)} & \textbf{Avg time (s)} \\
\midrule
\textbf{Visual Reasoning CAPTCHA}  & 100 & \textbf{31} & 3.55 \\
Text CAPTCHA              & 100 & 46 & 2.41 \\
Rotated CAPTCHA           & 100 & 45 & 4.38 \\
FunCAPTCHA                & 100 & 39 & 2.31 \\
Slider CAPTCHA            & 100 & 40 & 2.17 \\
3D CAPTCHA                & 100 & 54 & 2.03 \\
reCAPTCHA                 & 100 & 50 & 4.11 \\

\bottomrule
\end{tabularx}
\vspace{-0.1in}
\end{table}

\smallskip
\noindent\textbf{Empirical study on CAPTCHA types.} 
Table~\ref{tab:captcha_types_summary} reports zero-shot GPT-4o accuracy across major CAPTCHA types. While text-based, slider, and reCAPTCHA already expose clear weaknesses, VRCs stand out as the hardest: GPT-4o achieves only \textbf{31.2\%} accuracy. The failures stem from cluttered and noisy scenes, coupled with multi-constraint queries that require compositional reasoning. These highlight VRCs as the \emph{hardest frontier} in CAPTCHA design, resisting even SOTA LLMs.

\section{Experiment Supplementary}
\label{apd-exp}

\subsection{Dataset in Details} \label{sec:data_sources}
To evaluate the generalizability across diverse real-world CAPTCHA deployments, we curate a benchmark from six widely used VRC datasets. These datasets differ in visual composition, reasoning complexity, and task types, providing broad challenge characteristics (Table~\ref{tab:dataset_summary}). 

\smallskip
\noindent\textbf{VTT (Tencent).} 
Each VTT challenge consists of a natural language instruction and an image populated with 10 to 20 synthetic 3D objects, such as letters, shapes, and digits. VTT questions span three categories: attribute identification, comparative reasoning, and spatial reasoning, involving complex layouts and diverse object variations.

\begin{figure}[t]
    \centering
    \includegraphics[width=0.7\columnwidth]
    {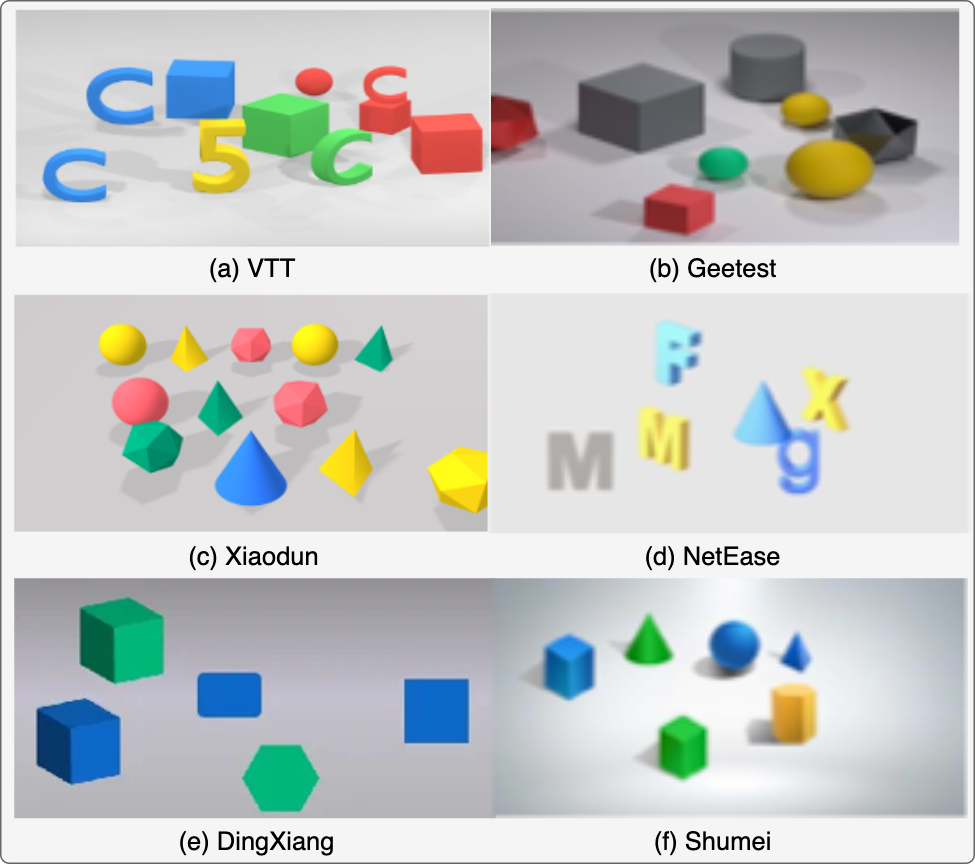}
    \caption{Six Widely Deployed VRC Datasets}
    \label{fig:datasets}
\end{figure}

\smallskip
\noindent\textbf{Geetest.}
Geetest challenges resemble VTT but use simpler compositions with 7–10 regular geometric objects. Prompts target shape, color, size, and relative location. Compared to VTT, they often introduce occlusion or ambiguous spatial relations, where targets are partially hidden or positioned unclearly. Distinct object and prompt formats further increase cross-domain variation.

\smallskip
\noindent\textbf{Xiaodun.}
Xiaodun’s space-reasoning CAPTCHA is inspired by VTT. Each challenge presents 12–14 3D objects (letters, numbers, shapes) and requires reasoning over shared attributes (e.g., \textit{same color as}'') and spatial relations (\textit{on top of}''). Unlike VTT, Xiaodun arranges objects densely and stacked, leading to frequent occlusions and contact-based relations that complicate detection.

\smallskip
\noindent\textbf{NetEase.}
The NetEase CAPTCHA dataset includes 5 to 7 regular geometric or alphanumeric objects per image. The prompts focus on orientation-based filters (e.g., ``\textit{side-facing}''), attribute consistency (e.g., ``\textit{same color}''), and basic directional reasoning. Compared to VTT and Xiaodun, the scene layout is cleaner, but questions still require object-level understanding.

\smallskip
\noindent\textbf{DingXiang.}
DingXiang CAPTCHA images typically display 5 planar or geometric objects and focus primarily on location-based reasoning, such as ``\textit{the object on the left}'', ``\textit{above}'', or ``\textit{closest to}''. No abstract logical reasoning or multi-stage dependencies are involved.

\smallskip
\noindent\textbf{Shumei.}
Shumei provides a more minimalistic style. Each image contains only 6 regular 3D geometric shapes. The prompt format is fixed: the user is always asked to click the smallest object with a specific color and shape.

\begin{figure}[htbp]
    \centering
    \includegraphics[width=0.5\linewidth]{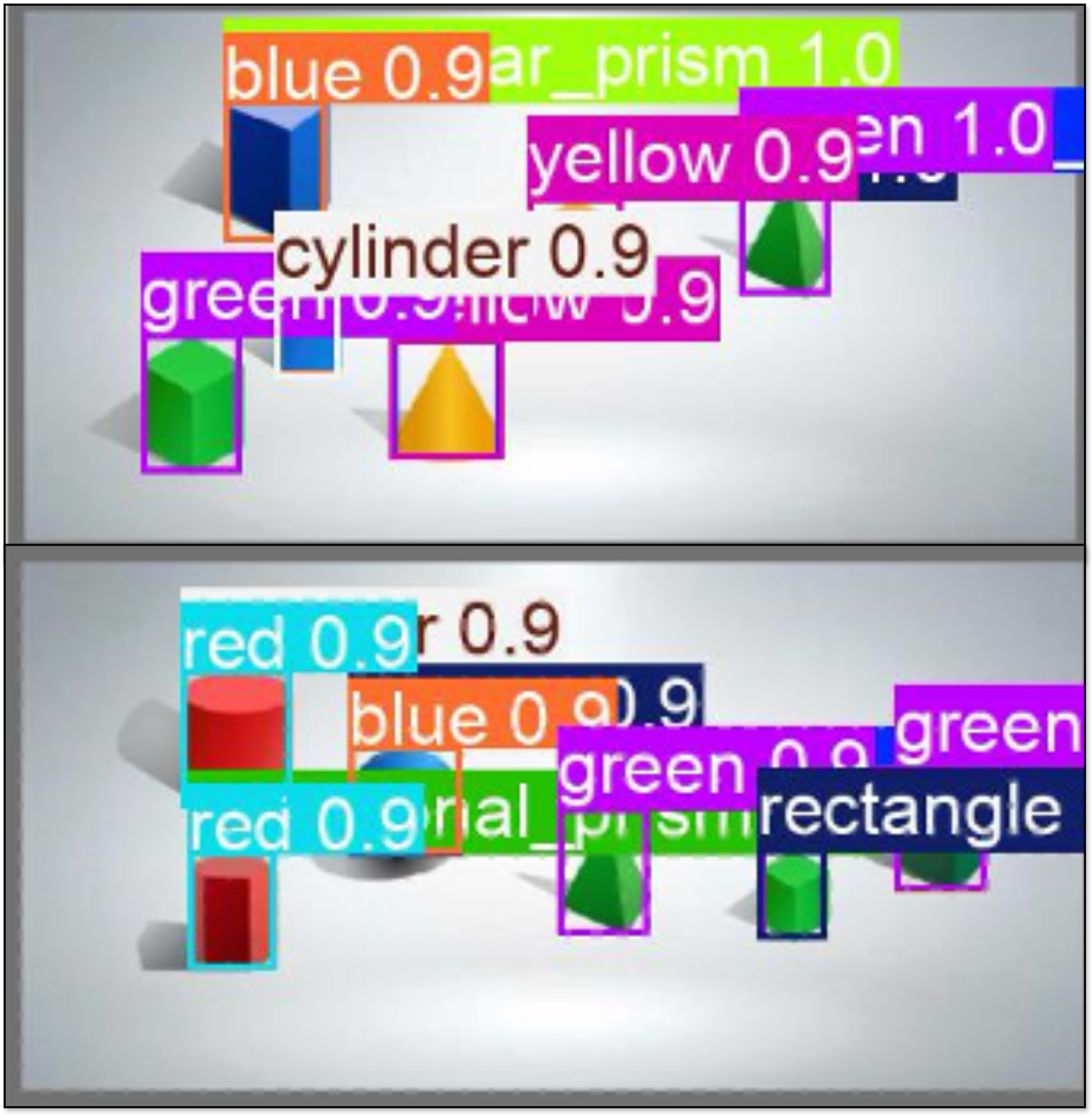}
    \caption{\textbf{Qualitative detection results.} Sample predictions of the detector. Each bounding box encodes the predicted compound label with its confidence score.}
    \label{fig:yolo_accuracy}
\end{figure}

\begin{figure*}[htbp]
    \centering
    \includegraphics[width=0.80\linewidth]{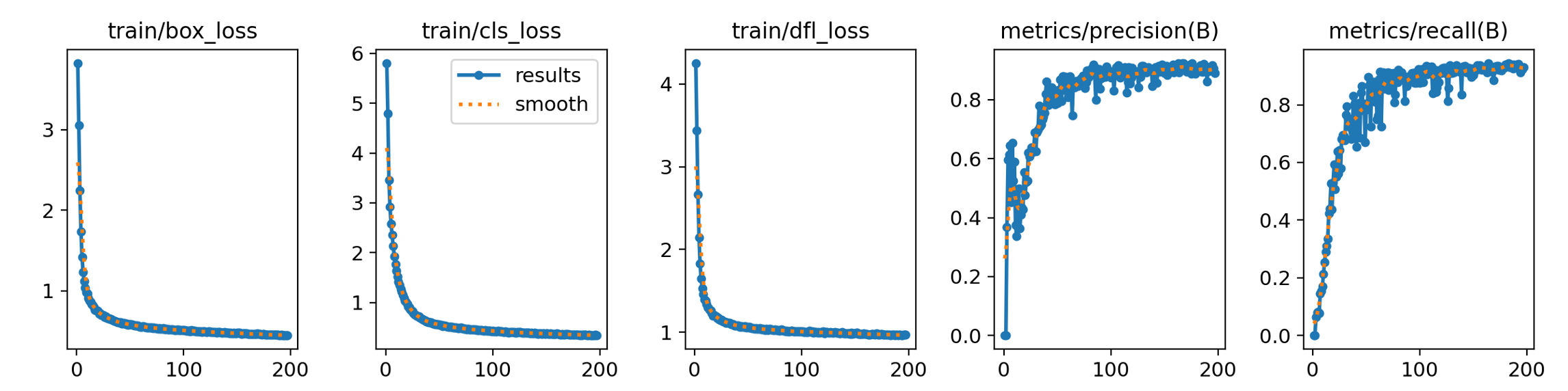}
    \caption{\textbf{Training and validation metrics of multi-object detection model.} 
    The plots illustrate the convergence across 200 epochs. 
    Metrics: training losses (\emph{box\_loss}, \emph{cls\_loss}, \emph{dfl\_loss}) and evaluation metrics (\emph{precision}, \emph{recall}). }
    \label{fig:train_results}
\end{figure*}

\subsection{Human Baseline}
\label{app:human}
We established a human baseline under the same benchmark setting used for all automated methods. Ten participants were recruited to solve a representative subset of the benchmark. Participants were university students and staff volunteers aged between 20 and 35. We note that this may limit the external validity of human baseline, and broader demographic coverage is left to future work.

For each VRC platforms, participants were presented with $100$ randomly sampled CAPTCHA challenges (a total of $1{,}000$ per platform), drawn uniformly from the identical evaluation corpus used in the model experiments. We recorded both correctness and response latency for every trial. Table~\ref{tab:human_baseline} summarizes the aggregated statistics.

\begin{table}[!htbp]
\centering
\scriptsize
\caption{\textbf{Human baseline (10 participants $\times$ 100 challenges per platform).} For each VRC platform we report the number of evaluated items ($N$), total correct answers, mean accuracy with across-participant standard deviation, mean response time with across-participant standard deviation, and the cumulative time across all 1{,}000 trials.}
\label{tab:human_baseline}
\vspace{7pt}
\begin{tabularx}{\columnwidth}{>{\columncolor{kimicolor}}c|c|cc|cc}
\toprule
\multicolumn{1}{c}{\textbf{Platform}} & \textbf{$N$} & \textbf{Correct} & \multicolumn{1}{c}{\textbf{Accuracy (\%)}} & \textbf{Mean Time (s)} & \textbf{Total Time} \\
\midrule
VTT       & 1000 & 876 & $87.60 \ (\pm 3.10)$ & $7.40 \ (\pm 2.10)$ & 02:03:20 \\
Geetest   & 1000 & 907 & $90.70 \ (\pm 2.70)$ & $7.80 \ (\pm 2.30)$ & 02:10:00 \\
NetEase   & 1000 & 951 & $95.12 \ (\pm 2.20)$ & $6.90 \ (\pm 1.90)$ & 01:55:00 \\
DingXiang & 1000 & 954 & $95.40 \ (\pm 2.00)$ & $6.60 \ (\pm 1.80)$ & 01:50:00 \\
Shumei    & 1000 & 932 & $93.20 \ (\pm 2.50)$ & $6.10 \ (\pm 1.70)$ & 01:41:40 \\
Xiaodun   & 1000 & 918 & $91.80 \ (\pm 2.80)$ & $7.20 \ (\pm 2.00)$ & 02:00:00 \\
\bottomrule
\end{tabularx}
\end{table}

\subsection{Multi-object Detector Performance}
\label{app:detector}


\smallskip
\noindent\textbf{Qualitative results.}  
In addition to quantitative metrics, we provide qualitative detection outcomes in Fig.~\ref{fig:yolo_accuracy}. The figure illustrates representative predictions produced by our multi-object detector across multiple VRC platforms. Each bounding box is annotated with the predicted compound label (capturing shape, color, and orientation attributes), together with its confidence score. The results show that the detector can reliably localize and classify diverse objects even under cluttered backgrounds and varying scales, confirming the robustness of the perception module in supporting downstream reasoning.

\smallskip
\noindent\textbf{Training dynamics.}  
Fig.~\ref{fig:train_results} shows that all training and validation losses (\emph{box}, \emph{cls}, \emph{dfl}) decrease sharply in the first 20 epochs and stabilize near zero, with validation curves following closely and exhibiting only minor variance. Precision and recall rise quickly and exceed 0.90 after 100 epochs, confirming both accurate localization and stable generalization. Smoothed curves (orange) align well with raw results, indicating low epoch-to-epoch variance and a robust optimization process.

\section{Case Study}

\smallskip
\noindent We illustrate a representative example of how \textsc{VIPER} processes a VRC (i.e., end-to-end perceptual-reasoning). This walkthrough highlights each module’s intermediate outputs, showing how raw images and queries are progressively transformed into a final answer. 

\smallskip
\noindent\textbf{Step 1: Input}. Image and query are given: 
\begin{center}
\fbox{%
\begin{minipage}{0.75\linewidth}
 \textit{``Please click on the object directly below the letter `T'.''}

\begin{center}
    \includegraphics[width=0.7\linewidth]{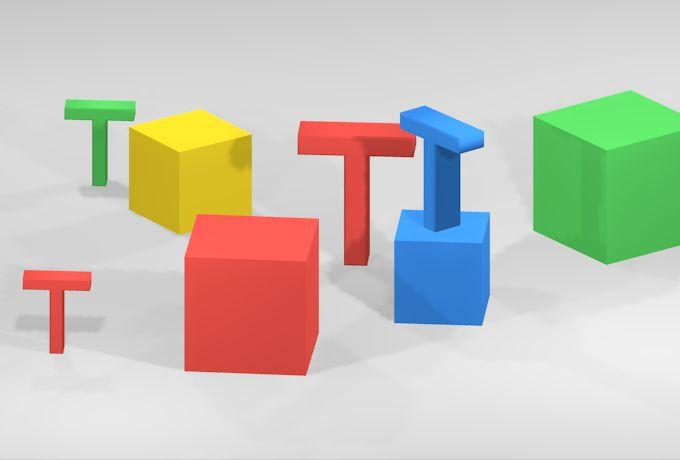}
\end{center}

\end{minipage}}
\vspace{5pt}
\end{center}

\smallskip
\noindent\textbf{Step 2: Perception output.}  
The detector processes the input image and produces a structured set of object candidates with their categories, coordinates, and bounding boxes. The complete output is shown below:

\begin{lstlisting}[language=json, basicstyle=\ttfamily\scriptsize, breaklines=true]
[
  {"Object": "T", "location": [442.5, 168.2],
   "bbox": [394.7731, 101.0462, 490.2843, 235.4473]},
  {"Object": "T", "location": [358.8, 192.3],
   "bbox": [292.8094, 114.6200, 424.7109, 270.0581]},
  {"Object": "blue", "location": [442.2, 267.8],
   "bbox": [388.4220, 203.7325, 496.0359, 331.7915]},
  {"Object": "T", "location": [56.1, 310.4],
   "bbox": [17.8713, 265.3660, 94.3498, 355.4301]},
   ...
  {"Object": "cube", "location": [442.2, 267.3],
   "bbox": [387.9681, 203.1617, 496.4134, 331.5254]},
  {"Object": "side", "location": [191.2, 171.7],
   "bbox": [123.6435, 106.6081, 258.7856, 236.8480]},
  {"Object": "red", "location": [252.0, 294.1],
   "bbox": [178.4103, 209.2727, 325.6867, 378.9393]},
  {"Object": "front", "location": [56.6, 310.4],
   "bbox": [19.0843, 265.4105, 94.0871, 355.3325]}
]
\end{lstlisting}

\smallskip
\noindent\textbf{Step 3: QIE result.}  
QIE processes the natural-language instruction $q$ and generates structured symbolic records for the mentioned objects. For this query, the output is:

\begin{lstlisting}[language=json, basicstyle=\ttfamily\scriptsize, breaklines=true]
[
  {"ObjectData": {"shape": "T", "color": "", "orientation": ""}}
]
\end{lstlisting}

\smallskip
\noindent\textbf{Step 4: Integrator output.}  
The integrator aligns the structured detections from the visual detector with the symbolic records produced by QIE. The resulting representation serves as a refined set of candidate objects for downstream relational reasoning. The output is:

\begin{lstlisting}[language=json, basicstyle=\ttfamily\scriptsize, breaklines=true]
[ {
    "Object": ["T", "blue", "side"],
    "location": [442.5, 168.2],
    "bbox": [394.7731, 101.0462, 490.2843, 235.4473]
  },
  {
    "Object": ["T", "front", "red"],
    "location": [358.8, 192.3],
    "bbox": [292.8094, 114.6200, 424.7109, 270.0581]
  },
  {
    "Object": ["T", "red", "front"],
    "location": [56.1, 310.4],
    "bbox": [17.8713, 265.3660, 94.3498, 355.4301]
  },
  {
    "Object": ["T", "green", "front"],
    "location": [99.3, 142.5],
    "bbox": [58.9634, 93.5350, 139.6709, 191.4095]
  }]
\end{lstlisting}

\smallskip
\noindent\textbf{Step 5: RPIE output.}  
Given detected reference objects (``\textit{T}''), RPIE shifts each center by a fixed offset $\Delta$ in a specified direction (``\textit{below}'') and selects detections whose boxes contain probe points, tagging ``\texttt{!maybe result}''.

\begin{lstlisting}[language=json, basicstyle=\ttfamily\scriptsize, breaklines=true]
[
  {"Object": ["T", "blue", "side"], "location": [442.5, 168.2]},
  {"Object": ["blue", "cube", "side", "!maybe result"], "location": [442.2, 267.8]},
  {"Object": ["T", "front", "red"], "location": [358.8, 192.3]},
  {"Object": ["T", "red", "front"], "location": [56.1, 310.4]},
  {"Object": ["T", "green", "front"], "location": [99.3, 142.5]}
]
\end{lstlisting}

\smallskip
\noindent\textbf{Step 6: Dynamic prompt.}  
The dynamic prompt explicitly instructs the LLM to translate the query, reason about relative spatial relations, and identify the correct target object from the candidates marked with \texttt{!maybe result}. The constructed prompt for this case is:

\begin{center}
\fbox{%
\begin{minipage}{0.90\linewidth}
\small

\textbf{Dynamic Prompt:}

You are a 3D reasoning captcha assistant. 
You will first receive the question of the captcha, 
then the captcha image, and the output produced by the 
visual model (which may contain errors). 

\textit{Important:} This is a question involving relative positional 
relations. Think carefully about the relationship between 
the reference object and the target object. Finally, provide 
the coordinates of the target object, not the coordinates 
of the reference object. 

\textit{Examples:}

- "Click the letter above the cylinder" means to click 
  the letter, not the cylinder.\\
- "Click the object below 'j'" means to click the object, 
  not the character 'j'.

Please select the answer from the candidates marked 
with "!maybe result".

\textit{Question:} "Click the object directly below the letter 'T'"

\end{minipage}}
\end{center}

\smallskip
\noindent\textbf{Step 7: LLM analyzer output.}  
The LLM receives the dynamic prompts and the candidate objects produced by the RPIE module. It reasons over the positional relation and selects the final answer. In this case, the output is:  

\begin{center}
\fbox{%
\begin{minipage}{0.90\linewidth}
\small

Based on the aligned detections, the the object directly below the letter "T" is the \textbf{blue cube}.  

The coordinates for the blue cube are \textbf{(442.2, 267.8)}.
\end{minipage}}
\end{center}

\smallskip
\noindent\textbf{Step 8: Verification.}  
The system checks whether the coordinates predicted by the LLM Analyzer fall inside the ground-truth bounding box of the target region. 

\begin{center}
\fbox{%
\begin{minipage}{0.90\linewidth}
\small

The annotated target bounding box is defined as:  
Left = 0.5670, Right = 0.7352, Top = 0.4469, Bottom = 0.7288.  

The normalized point (0.6503, 0.5822) lies inside this region.  
\textbf{Result: Successful verification.}

\begin{center}
    \includegraphics[width=0.75\linewidth]{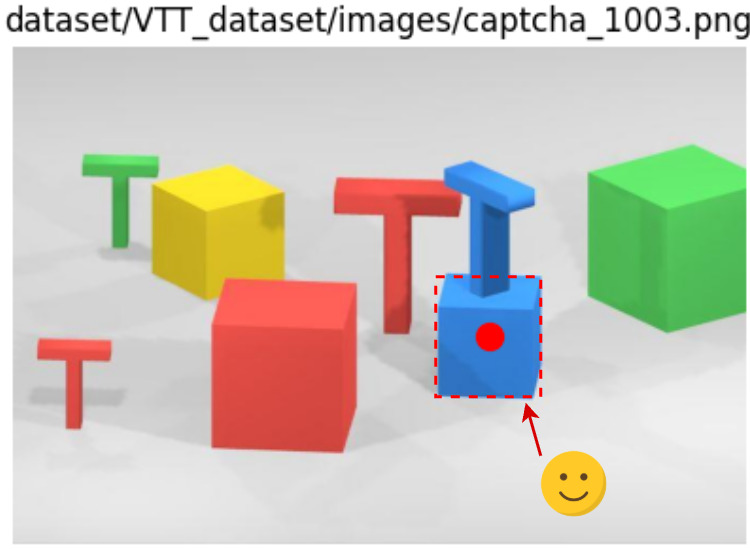}
\end{center}

\end{minipage}}
\end{center}

\end{document}